\documentclass[12pt]{article}
\pdfoutput=1

\usepackage[koi8-r]{inputenc}
\usepackage[english]{babel}

\usepackage{url}
\usepackage{color}
\usepackage{amsmath}
\usepackage{epsf}
\usepackage{epsfig}
\usepackage{here}
\usepackage{amssymb}
\usepackage{cite}
\usepackage{graphicx}
\def\e{{\epsilon}}

\newcommand{\ba}{\begin{array}}
\newcommand{\ea}{\end{array}}
\newcommand{\be}{\begin{equation}}
\newcommand{\ee}{\end{equation}}
\newcommand{\bea}{\begin{eqnarray}}
\newcommand{\eea}{\end{eqnarray}}
\newcommand{\bg}{\begin{gather}}
\newcommand{\eg}{\end{gather}}
\newcommand{\bseq}{\begin{subequations}}
\newcommand{\eseq}{\end{subequations}}

\makeatletter

\def\lsim{\compoundrel<\over\sim}
\def\compoundrel#1\over#2{\mathpalette\compoundreL{{#1}\over{#2}}}
\def\compoundreL#1#2{\compoundREL#1#2}
\def\compoundREL#1#2\over#3{\mathrel
         {\vcenter{\hbox{$\m@th\buildrel{#1#2}\over{#1#3}$}}}}
\makeatother

\begin{document}

\begin{center}

    {\Large\bf Bounds on non-standard interactions of neutrinos from IceCube DeepCore data}
    \\
    \vspace{0.8cm}
    \vspace{0.3cm}
    S.~V.~Demidov$^{a,}$\footnote{{\bf e-mail}: demidov@ms2.inr.ac.ru}, 
    \\
    
    $^a${\small{\em 
        Institute for Nuclear Research of the Russian Academy of Sciences, }}\\
      {\small{\em
          60th October Anniversary prospect 7a, Moscow 117312, Russia
      }
      }
%%       \\
%% $^{b}${\small{\em
%% Moscow Institute of Physics and Technology,
%% }}\\
%% {\small{\em
%% Institutsky per. 9, 
%%   Dolgoprudny 141700, Russia
%% }}\\
  \end{center}
 % \vspace{2cm}
\begin{abstract}
  New physics in neutrino sector can reveal itself via non-standard
  neutrino interactions which can result in modification of
  the standard picture of neutrino propagation in matter. Experiments with 
  atmospheric neutrinos has been used to probe this scenario. Using
  publicly available three-year low energy data in IceCube DeepCore we
  place bounds on the parameters $\epsilon_{\alpha\beta}$ of
  non-standard neutrino interactions in propagation. We obtain
  restrictive   constraints not only for $\mu\tau$ sector but also 
  for flavor changing interactions involving electron neutrinos.
\end{abstract}

\section{Introduction}

Phenomenon of neutrino oscillations is well established by many
experiments~\cite{Tanabashi:2018oca}. It implies nontrivial mixing
between neutrinos and non-zero values of their masses. Explanation of
these properties of neutrinos lies beyond the Standard Model and
requires new physics. The latter may reveal itself via
non-renormalizable interactions involving neutrino fields. In
particular, new Fermi-type interactions of neutrinos with SM fermions
are of interest and they are known as non-standard neutrino
interactions~\cite{Wolfenstein:1977ue,Guzzo:1991hi}. In what follows we consider 
neutral current (NC) or matter non-standard neutrino interactions (NSI)
with the lagrangian 
\be
\label{eq:1:1}
   {\cal L}^{NC}_{NSI} =
   -\sum_{f, P=P_L,P_R} \epsilon_{\alpha\beta}^{fP} 2\sqrt{2}G_F
   (\bar{\nu}_{\alpha}\gamma^{\mu}P_L\nu_\beta)  
   (\bar{f}\gamma_{\mu}Pf).
\ee
Here $P_{L,R}$ are the chirality projectors, $\epsilon_{\alpha\beta}^{fP}$
are the NSI parameters and sum goes over all SM
fermions~$f$. Note that the lagrangian~\eqref{eq:1:1} contains only
operators which do not change flavor of the fermion~$f$. The flavor
changing interactions  of the type~\eqref{eq:1:1} 
are severely constrained from results on lepton flavor violating and
FCNC processes. NC NSI~\eqref{eq:1:1} can modify neutrino phenomenology
in several  ways and these interactions can reveal themselves in
scattering processes as well as in neutrino oscillation experiments
(see
Refs.~\cite{Ohlsson:2012kf,Miranda:2015dra,Farzan:2017xzy,Dev:2019anc}  
for reviews). Examples of phenomenologically viable models
  predicting the matter NSI with sizable values of couplings were
  discussed e.g. in
  Refs.~\cite{Farzan:2015doa,Farzan:2015hkd,Farzan:2016wym,Farzan:2016fmy,Babu:2019mfe}.

One of the consequences of the interactions~\eqref{eq:1:1} is
modification of neutrino propagation in matter. In the presence of NSI
the evolution of neutrino having energy $E$ is described by the
Hamiltonian 
\be
\label{eq:1:2}
H = \frac{1}{2E}U{\rm diag}(0,\Delta m_{21}^2,\Delta m_{31}^2)U^{\dagger}
+ V_e\epsilon^{m},
\ee
where $U$ is the vacuum Pontecorvo-Maki-Nakagawa-Sakata (PMNS) matrix
and $\Delta m_{21}^2, \Delta m_{31}^2$ are differences of the neutrino
masses squared. The last term in~\eqref{eq:1:2} describes matter
effects and it depends on the matter density through $V_e =
(-)\sqrt{2}G_FN_e$ for (anti)neutrinos,
where $N_e$ is the electron number density. The NSI parameters from
the interaction lagrangian
\eqref{eq:1:1} enter the Hamiltonian as follows
\be
\label{eq:1:3}
\e^m = 
\left(
\begin{array}{ccc}
1+\e_{ee} & \e_{e\mu} & \e_{e\tau} \\
\e_{e\mu}^*       & \e_{\mu\mu} & \e_{\mu\tau} \\
\e_{e\tau}^*       &  \e_{\mu\tau}^*          &  \e_{\tau\tau}
\end{array}
\right),
\ee
where 
\be
\label{eq:1:4}
\e_{\alpha\beta} = \sum_{f=e,u,d}(\e_{\alpha\beta}^{f P_L} +
\e_{\alpha\beta}^{f P_R})\frac{N_f}{N_e}\,.
\ee
Here $N_f$ is the number density of the fermion $f$ in matter. In this study we
concentrate on neutrino propagation in the Earth and in this case the
expression~\eqref{eq:1:4} for NSI parameters transforms into
\be
\label{eq:1:5}
\epsilon_{\alpha\beta} \approx
\epsilon_{\alpha\beta}^{eV}+3\frac{N_u}{N_e}\epsilon_{\alpha\beta}^{uV}
+ 3\frac{N_d}{N_e}\epsilon_{\alpha\beta}^{dV}\,,
\ee
where $\epsilon_{\alpha\beta}^{fV}\equiv
\epsilon_{\alpha\beta}^{fP_{L}}+\epsilon_{\alpha\beta}^{fP_R}$. In
what follows the notation $\epsilon_{\alpha\beta}$ refers
to~Eq.\eqref{eq:1:5}. In general the parameters
$\epsilon_{\alpha\beta}$ are complex-valued numbers. Here we take them
real for simplicity (see~\cite{Esteban:2019lfo} for recent discussion 
of the effect of CP violation in NSI). 

Atmospheric neutrinos is an important tool to explore neutrino
properties 
and, in particular, to search for new interactions in neutrino
sector. The impact 
of NSI in propagation at experiments with atmospheric neutrinos has
been studied extensively
(see~\cite{GonzalezGarcia:2004wg,Friedland:2004ah,Blennow:2005qj,Friedland:2005vy,GonzalezGarcia:2005xw,Mann:2010jz,Ohlsson:2013epa,Choubey:2014iia,GonzalezGarcia:2011my,Mocioiu:2014gua,Chatterjee:2014gxa,Fukasawa:2016nwn}
for an incomplete list). Probes 
with atmospheric neutrinos has been used to  
constrain the matter NSI in the Earth. Results  
of experiments with atmospheric neutrinos allow to put rather
stringent bounds~\cite{Mitsuka:2011ty,Esmaili:2013fva,Fukasawa:2015jaa,  
  Salvado:2016uqu,Aartsen:2017xtt} on the parameters
$\epsilon_{\alpha\beta}$. 
 Earlier studies of NSI with IceCube
  data~\cite{Esmaili:2013fva,Salvado:2016uqu, Aartsen:2017xtt} put
  constraints on the NSI parameters in $\mu\tau$
  sector. Preliminary results with the bounds on more generic NSI
  models were recently reported
  in~\cite{ic_poster,ic_talk1,ic_talk2} by IceCube collaboration.  In the present study we use the 
publicly available IceCube DeepCore three-year low energy data
sample~\cite{ic_data} and perform an independent analysis to constrain
the parameters $\epsilon_{\alpha\beta}$. This data sample is
very close to what was used by IceCube to measure the neutrino
oscillation parameters in  Ref.~\cite{Aartsen:2017nmd}. We perform an
analysis of the NSI effect on atmospheric neutrino propagation in the
Earth. Information provided by IceCube with the data
release~\cite{ic_data} and, in particular, results of Monte-Carlo
simulation, expected background from atmospheric muons as well as
parametrization of estimated instrumental systematic effects, allows
one to made a realistic prediction for  number of expected events in
models with non-zero matter NSI parameters and compare them against
the data. As a result we obtain allowed regions for the parameters 
$\epsilon_{\alpha\beta}$ under certain model assumptions. 
 We compare them with results from other oscillation experiments
  and discuss impact of systematic uncertainties.

The rest of the paper is organized as follows.
In Section~2 we describe methodology used to bound
$\epsilon_{\alpha\beta}$ with the IceCube DeepCore three-year data
sample. In Section~3 we present our results. Section~4 is reserved
for conclusions. 

\section{Description of the analysis}
For the present study we use publicly available three-year data
sample~\cite{ic_data} in IceCube DeepCore which is referred to as
'Sample B' in\footnote{Another data sample in this
  release referred to as 'Sample A' was used for
  measurement of atmospheric tau neutrino
  appearance~\cite{Aartsen:2019tjl} and probing for neutrino mass
  ordering~\cite{Aartsen:2019eht}.} Ref.~\cite{Aartsen:2019tjl} and which is very close to
what was used in the oscillation analysis~\cite{Aartsen:2017nmd}. This
sample contains 40920 events in total which are distributed over
$8\times8\times2$ binned histogram. The latter is composed of equally
spaced bins in $\log_{10}{E^{reco}}\in[0.75,1.75]$, 8 equally spaced
bins in $\cos{\theta^{reco}}\in[-1,1]$ as well as 2 bins which
correspond to the track-like and cascade-like events. Here $E^{reco}$
and 
$\theta^{reco}$ refer to reconstructed values of neutrino energy and
zenith angle.  We use the public IceCube Monte-Carlo provided along
with the data sample to model the detector response and to relate
physical 
values of the energy, zenith angle and type of neutrino with
reconstructed characteristics of the events. The simulated neutrino
sample released 
at~\cite{ic_data} allows to calculate for each bin $i$ the effective areas
$A^{\nu_\alpha, t}_{i}(E,\cos{\theta})$ to be converted with predicted
neutrino flux to obtain expected number of events in the bins. The
effective areas are obtained as functions of true neutrino energy $E$
in the range from 1 to 1000~GeV, zenith angle $\theta$, neutrino type 
$\nu_\alpha$ and type of neutrino interaction $t$ (CC or NC).

For prediction of the neutrino flux at the detector level we start
with the atmospheric neutrino fluxes
$\Phi^{atm}_{\nu_\alpha}(E,\theta)$ 
for $\nu_e,\bar{\nu}_e,\nu_\mu$ and
$\bar{\nu}_\mu$ taken from Ref.~\cite{Honda:2015fha}. To describe
propagation of the atmospheric neutrinos in the Earth  in presence of
NSI 
one should solve the Schrodinger equation with the
Hamiltonian~\eqref{eq:1:2} for the case of varying density. In the
present analysis we solve it numerically as described
in~\cite{Boliev:2013ai,Demidov:2017mmw} using the algorithm presented
in~\cite{Ohlsson:1999xb}. For calculation of the electron number
density in the Earth we use PREM~\cite{Dziewonski:1981xy}. We
fix the following values\footnote{We note that analysis of solar
  neutrino propagation   revealed~\cite{Miranda:2004nb} that models
  with NSI allow for so-called LMA-D solution for the oscillation
  parameters for which $\sin^{2}{\theta_{12}}>0.5$. In this study we
  do not consider this possibility because, on the one hand, the
  impact of $\sin^2{\theta_{12}}$ on the results of our analysis is
  very mild and, on the other, recent studies (see
  e.g.~\cite{Coloma:2017ncl,Giunti:2019xpr,Coloma:2019mbs}) showed 
  that the LMA-D scenario is disfavored by experimental data.}
 of neutrino oscillation parameters:
$\sin^2{\theta_{12}}=0.304$, $\sin^2{\theta_{13}}=0.0217$ and $\Delta 
m^2_{21}=7.53\cdot 10^{-5}$~eV$^2$ and set $\delta_{CP}=0$. Upon
obtaining solution to the Schrodinger equation we calculate the
transition probabilities $P_{\nu_\alpha\to\nu_\beta}(E,\cos{\theta})$
to find neutrino of a flavor $\nu_\beta$ at the detector level from
neutrino of a flavor $\nu_\alpha$ produced in the 
atmosphere. It is the probability functions
$P_{\nu_\alpha\to\nu_\beta}(E,\cos{\theta})$ which 
depend on the matter NSI parameters $\epsilon_{\alpha\beta}$. The
resulting neutrino fluxes at the detector level are then obtained as 
\be
\label{eq:2:0}
\Phi^{det}_{\nu_\beta}(E,\cos{\theta}) =
\sum_{\nu_\alpha}\Phi^{atm}_{\nu_\alpha}(E,\cos{\theta})P_{\nu_\alpha\to\nu_\beta}(E,\cos{\theta})\,.
\ee 
and the expected number of events in $i$-th bin can be found as
follows
\be
\label{eq:2:00}
n_i^{\nu} = T\cdot \sum_{t, \nu_\alpha}\int\, dE\,d\cos{\theta}\,
A^{\nu_\alpha,t}_{i}(E,\cos{\theta})\Phi^{det}_{\nu_\alpha}(E,\cos{\theta})\,, 
\ee
where $T$ is the lifetime for the data sample under consideration
and sum goes over contributions from different neutrino flavors
$\nu_\alpha$ (including antineutrinos) and interaction types $t$ (CC
or NC). 

To compare the expected event distribution with the IceCube DeepCore
data we take into account systematic uncertainties and our
procedure includes a set of relevant nuisance parameters
$\eta_j$. They correspond to overall normalization of atmospheric
neutrino flux with no prior, the spectral index of the atmospheric
neutrino flux with the nominal value $\gamma=-2.66$ and  
a prior $\sigma_\gamma=0.1$, relative normalizations of $\nu_e,
\bar{\nu}_e$ events and NC events with an uncertainty 20\%. Also we
include additional corrections to the atmospheric neutrino flux at
production to take 
into account uncertainties in the hadron production in
atmosphere. This corrections as functions of neutrino energy and
zenith angle 
have been chosen to reproduce the uncertainties estimated
in Ref.~\cite{Barr:2006it} similar to how it was done
in~\cite{Terliuk:2018xom}. Also we take into account uncertainty
related to the neutrino nucleon cross section with baryon resonance 
production which is important for low energy part of the neutrino
sample. For that we introduce an additional nuisance parameter for
the contribution from such type of the events with 40\% 
uncertainty which is close to what was found
in~\cite{Terliuk:2018xom}. As discussed in~\cite{Aartsen:2017nmd}
uncertainties on the DIS cross section have negligible impact on the
results. We include in the analysis the template for 
the background of atmospheric muons with corresponding uncorrelated
error $\sigma^{uncor}_{\nu,\mu_{atm}}$ provided by IceCube for this
data sample~\cite{ic_data}. Normalization of the background is taken
as a nuisance 
parameter with no prior. Finally, we account for 
instrumental systematic uncertainties related to the optical
efficiencies of DOMs and relevant properties of the ice. These
uncertainties are included in our analysis as described
in~\cite{ic_data}. In summary, our implementation of the systematic
uncertainties is very close to how it was done in the original
oscillation analysis~\cite{Aartsen:2017nmd}.  We will discuss the
  impact of the most important systematic errors in the next Section.

Let us note, that in general interactions of the type~\eqref{eq:1:1}
result in changes of the NC neutrino-nucleon cross section and thus
can modify expected number of events. However, this effect is very
model dependent. Not only it depends on other combinations of the
parameters $\epsilon^{fP}_{\alpha\beta}$ than those in
Eq.~\eqref{eq:1:5}, but it is also affected by microscopic model
behind the effective 
lagrangian~\eqref{eq:1:1} and the resulting cross section will be
different fore models with light and heavy mediators
In the present analysis we
conservatively do not take 
into account the impact of NSI on the modification on the NC
neutrino-nucleon cross sections. 

To obtain bounds on the parameters of non-standard neutrino
interactions we define (c.f. Eq.(2) in~\cite{Aartsen:2017nmd}) 
\be
\label{eq:2:1}
\chi^2 =
\sum_i\frac{\left(n_i^\nu+n_i^{\mu_{atm}}-n_i^{data}\right)^2}
    {\left(\sigma_i^{data}\right)^2+ 
  \left(\sigma_{\nu,\mu_{atm},i}^{uncor}\right)^2} + \sum_j
\frac{\left(\eta_j-\hat{\eta}_j\right)^2}{\sigma_{\eta_j}^2}\,,  
\ee
where $n_i^\nu$ ($n_i^{\mu_{atm}}$) is the expected number of events
from atmospheric neutrinos (muons), $n_i^{data}$ is the
number of data events in the $i$-th bin,
$\sigma_i^{data}=\sqrt{n^{\nu}_{i}}$, and the first sum goes over all 
bins in the data sample. The second term accounts for contribution from 
the nuisance parameters $\eta_j$, given their default values
$\hat{\eta}_j$ and uncertainties $\sigma_{\eta_j}$. We fix  
$\sin^2{\theta_{12}}$, $\sin^2{\theta_{13}}$ and $\Delta
m^2_{21}$ as described above. We checked that the uncertainties in
their values produce negligible effect on the final results.
Also we assume $\delta_{CP}=0$. Other neutrino oscillation
parameters, $\sin^2{\theta_{23}}$ and $\Delta m^2_{31}$, are not
fixed to any {\it a priori} value but were determined from the
analysis itself in most part of the study. As a 
consistency check we reproduce confidence regions for
$\sin^2{\theta_{23}}$ and $\Delta m^2_{31}$ from the IceCube DeepCore
data sample assuming $\epsilon_{\alpha\beta}=0$. In particular, we
find for the case of normal  
mass ordering $\sin^2{\theta_{23}}=0.52^{+0.07}_{-0.08}$, $\Delta
m_{32}^2=2.29^{+0.16}_{-0.15}$~eV$^2$ which is very close to those
intervals 
obtained in Ref.~\cite{Aartsen:2017nmd}. In our analysis with
non-vanishing matter NSI parameters we consider $\sin^2{\theta_{23}}$
and $\Delta m^2_{31}$ as nuisance parameters with no prior  unless
stated otherwise. In  Section~3 we discuss how the bounds on
$\epsilon_{\alpha\beta}$ are robust if $\sin^2{\theta_{23}}$ and
$\Delta m_{31}^2$ are fixed in the analysis.

\begin{figure}[!htb]
\begin{picture}(300,200)(0,30)
\put(30,240){\includegraphics[angle=-90,width=0.40\textwidth]{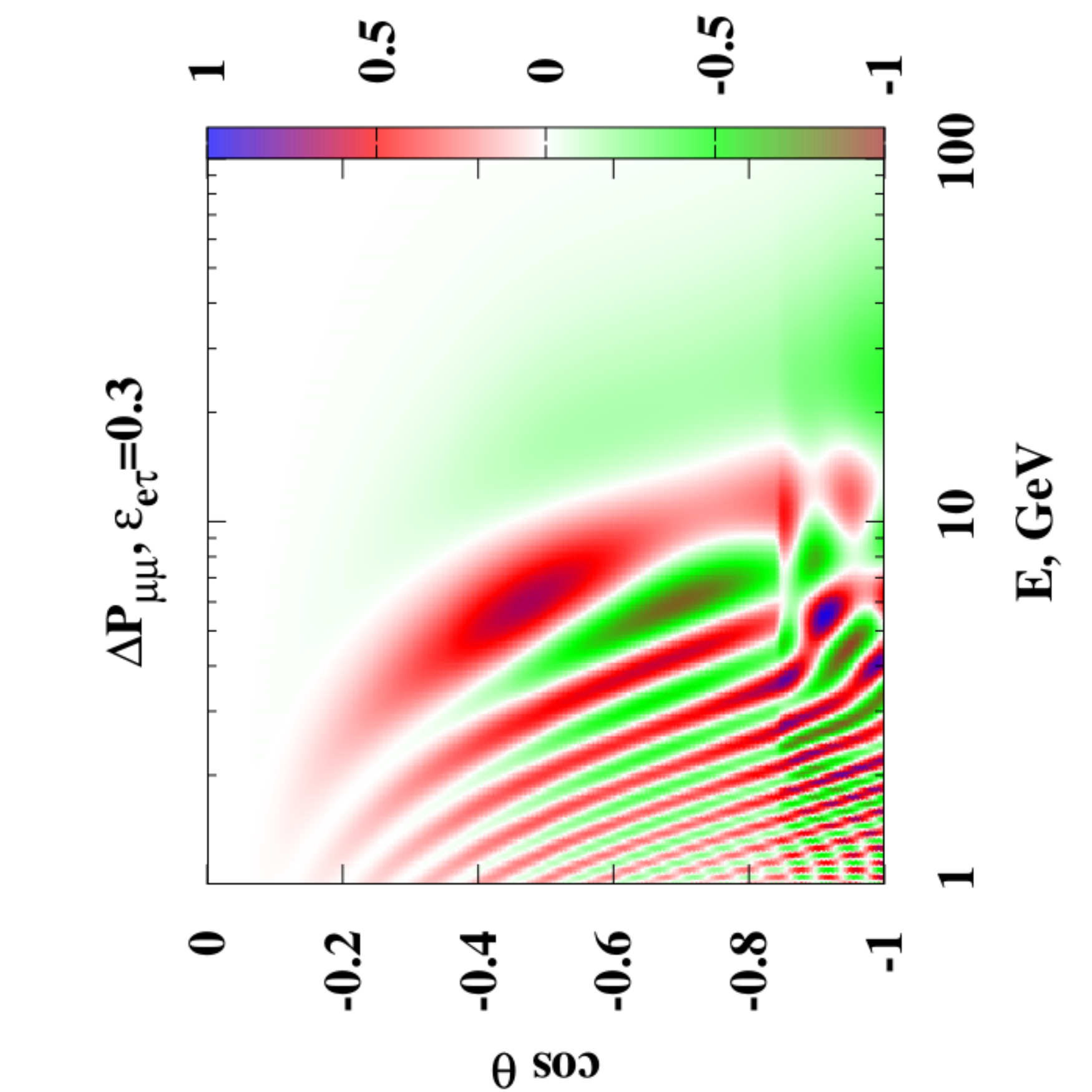}}
\put(250,240){\includegraphics[angle=-90,width=0.40\textwidth]{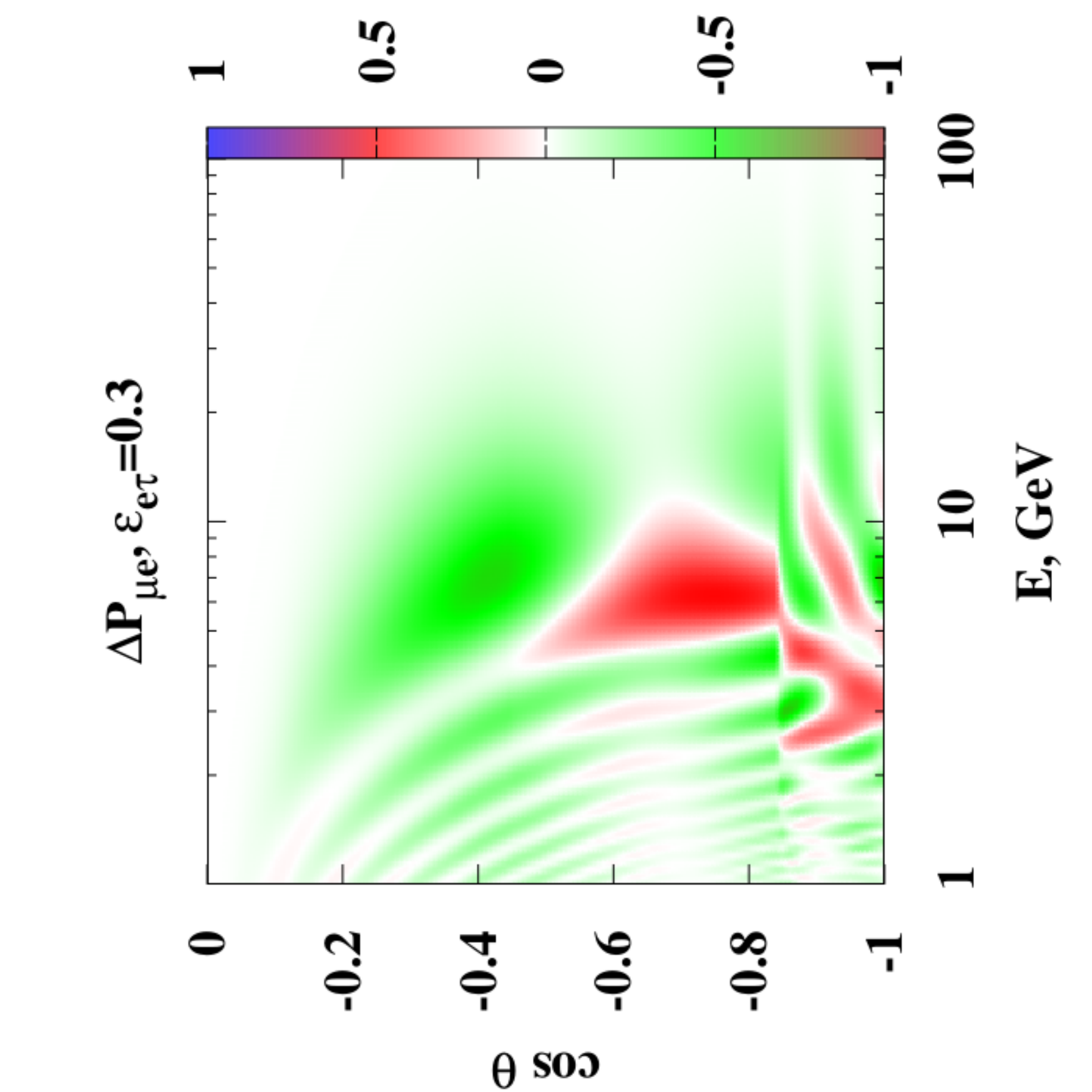}}
\end{picture}
\caption{\label{etau_osc} Differences between probabilities $\Delta
  P_{\mu\mu}\equiv
  P_{\nu_\mu\to\nu_\mu}^{NSI}-P_{\nu_\mu\to\nu_\mu}^{noNSI}$ (left
  panel) and $\Delta P_{\mu e}\equiv
  P_{\nu_\mu\to\nu_e}^{NSI}-P_{\nu_\mu\to\nu_e}^{noNSI}$ (right panel)
  shown in color as functions of $E$ and $\cos{\theta}$. The
  probabilities are calculated with $\epsilon^{e\tau}=0.3$ assuming
  $\sin^2{\theta_{23}}=0.51$ and $\Delta m_{31}^2 = 2.5\times
  10^{-3}$~eV$^2$. } 
\end{figure}

\begin{figure}[!htb]
\begin{picture}(300,200)(0,30)
\put(30,240){\includegraphics[angle=-90,width=0.40\textwidth]{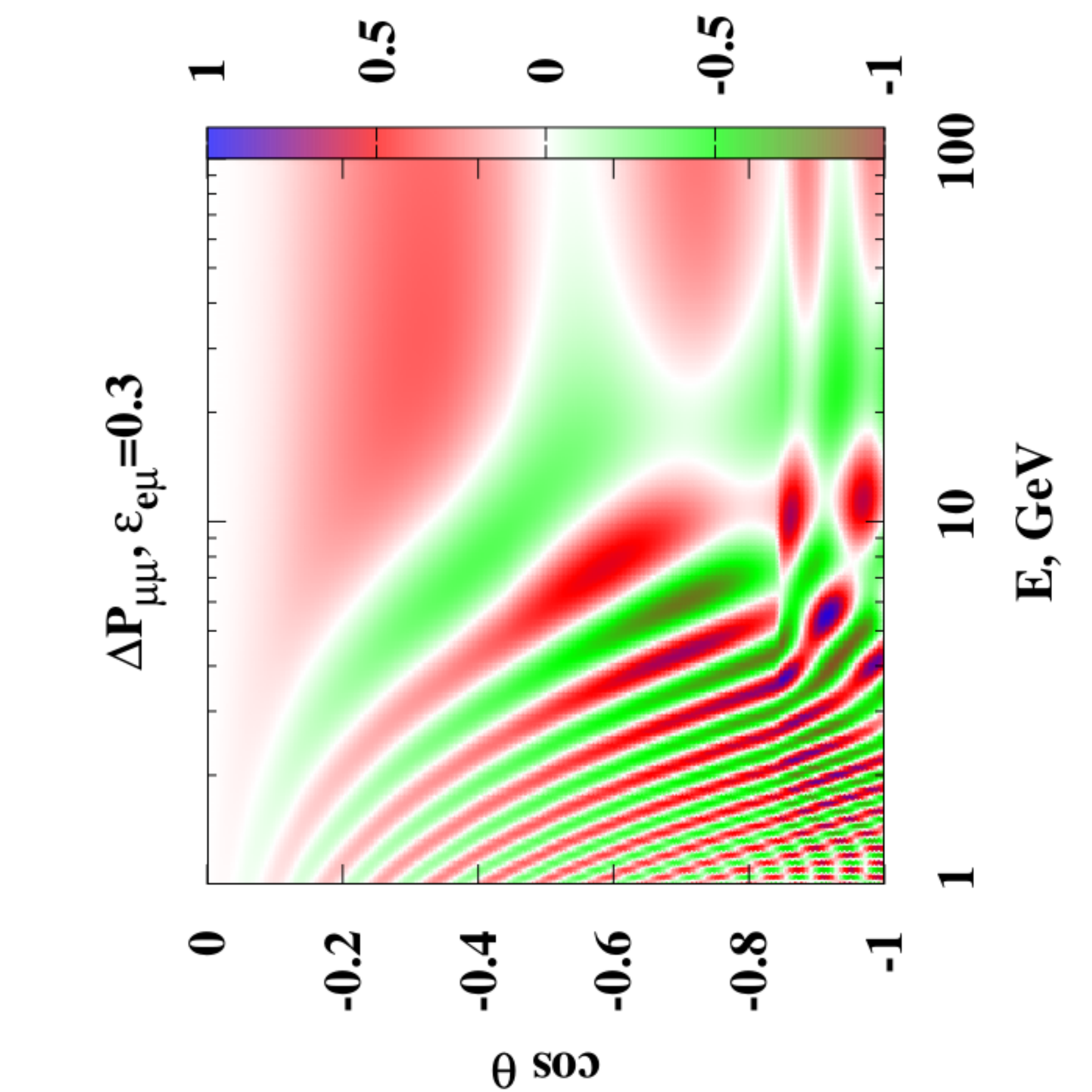}}
\put(250,240){\includegraphics[angle=-90,width=0.40\textwidth]{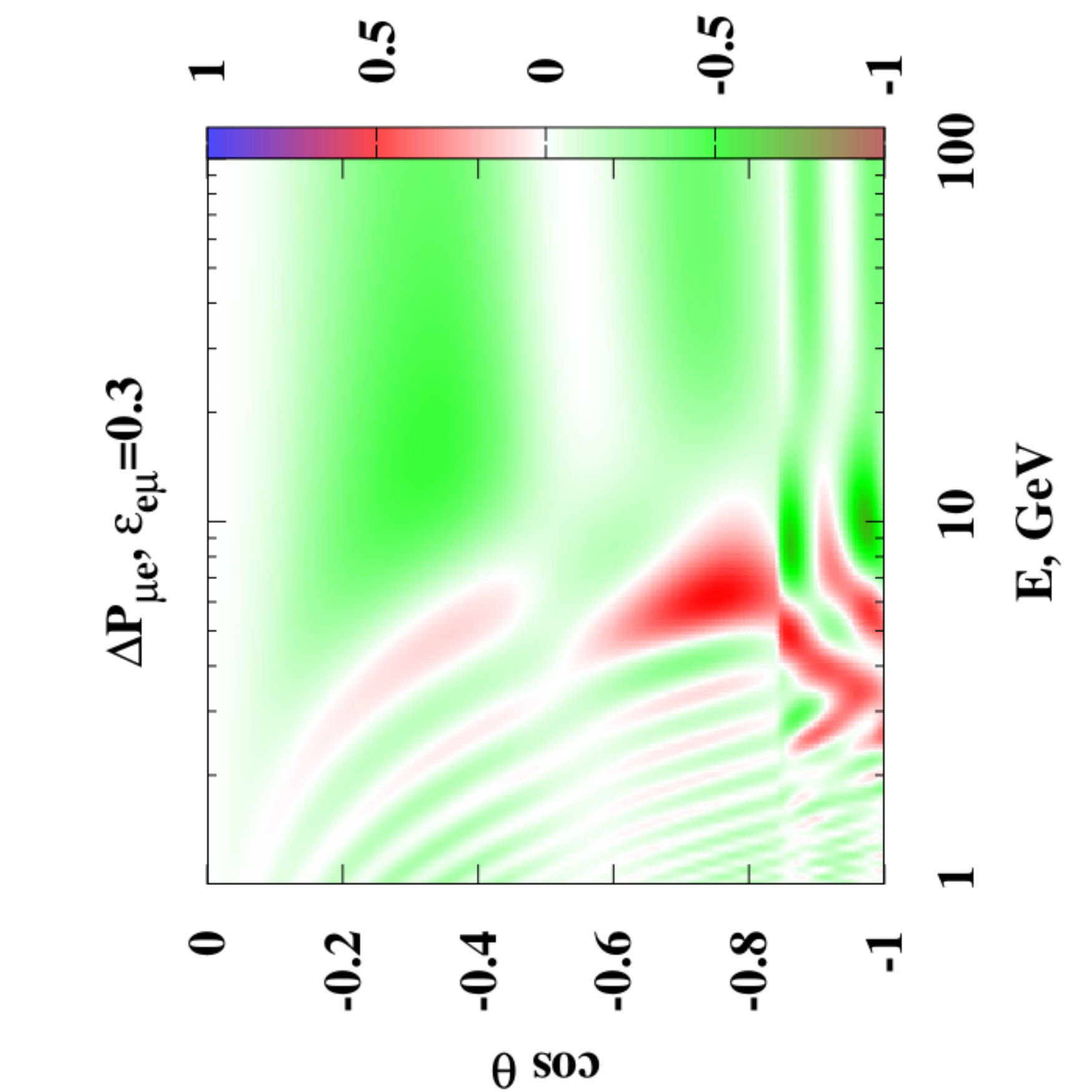}}
\end{picture}
\caption{\label{emu_osc} The same as in Fig.~\ref{etau_osc} but with
  $\epsilon_{e\mu}=0.3$.}  
\end{figure}

The IceCube DeepCore low energy data sample of the atmospheric 
neutrinos~\cite{ic_data} 
contains (contrary to the earlier IceCube DeepCore oscillation
analyses~\cite{Aartsen:2014yll}) not only track-like but also
cascade-like events which 
results mainly from CC interactions of $\nu_\mu$ and $\nu_e$ (see
Fig.~1 in Ref.~\cite{Aartsen:2017nmd}). This makes the data sample
sensitive not only to changes in the muon neutrino flux but also to
modification in the flux of electron neutrinos. Most of the
theoretical studies discuss impact of the NSI in propagation mainly on
the muon neutrino flux $\Phi^{det}_{\nu_\mu}$. This is mostly
sufficient for the models with 
non-zero values of the corresponding parameters in $\mu\tau$ sector
where the NSI effect on the electron neutrino flux is very mild given
current bounds on the NSI parameters. However, this is not the case
for models with non-zero $\epsilon_{e\tau}$ or $\epsilon_{e\mu}$. For
illustration, in Figs.~\ref{etau_osc} and~\ref{emu_osc}
we show oscillograms for $\Delta P_{\mu\mu}$ and $\Delta P_{\mu e}$
which are the differences between the transition probabilities with
and without NSI, i.e. $\Delta P_{\alpha\beta} =
P_{\nu_\alpha\to\nu_\beta}^{NSI} -
P_{\nu_\alpha\to\nu_\beta}^{noNSI}$ for neutrinos. Fig.~\ref{etau_osc}
and~\ref{emu_osc} correspond to $\epsilon_{e\tau}=0.3$ 
and $\epsilon_{e\mu}=0.3$, respectively. Right panels on both Figures correspond to
$\Delta P_{\mu\mu}$ while left are reserved for $\Delta P_{\mu e}$. We
observe 
that the off-diagonal matter NSI involving electron neutrinos may
considerably modify not only the fluxes of muon neutrinos but also
that of 
electron neutrinos. Note that the effect of non-zero
$\epsilon_{e\tau}$ decreases with increase of neutrino energy but
this is not the 
case for $\epsilon_{e\mu}$. Thus we expect that IceCube data for
neutrinos at high energies will be also sensitive to the parameter 
$\epsilon_{e\mu}$. 

\section{Results}
In this Section we present results on the allowed regions for
several matter NSI parameters $\epsilon_{\alpha\beta}$ from the
analysis of the low energy three-year IceCube DeepCore data
sample~\cite{ic_data}.  In this study we limit ourselves to a
  constrained analysis in the matter NSI parameter space assuming only
some of $\epsilon_{\alpha\beta}$ to be non-zero.
As neutrino oscillation probabilities depend on
differences between the
diagonal elements $\epsilon_{ee}-\epsilon_{\mu\mu}$ and
$\epsilon_{\tau\tau}-\epsilon_{\mu\mu}$ we fix $\epsilon_{\mu\mu} =
0$ in what follows.

\begin{figure}[t]
\begin{picture}(300,130)(0,20)
\put(30,140){\includegraphics[angle=-90,width=0.40\textwidth]{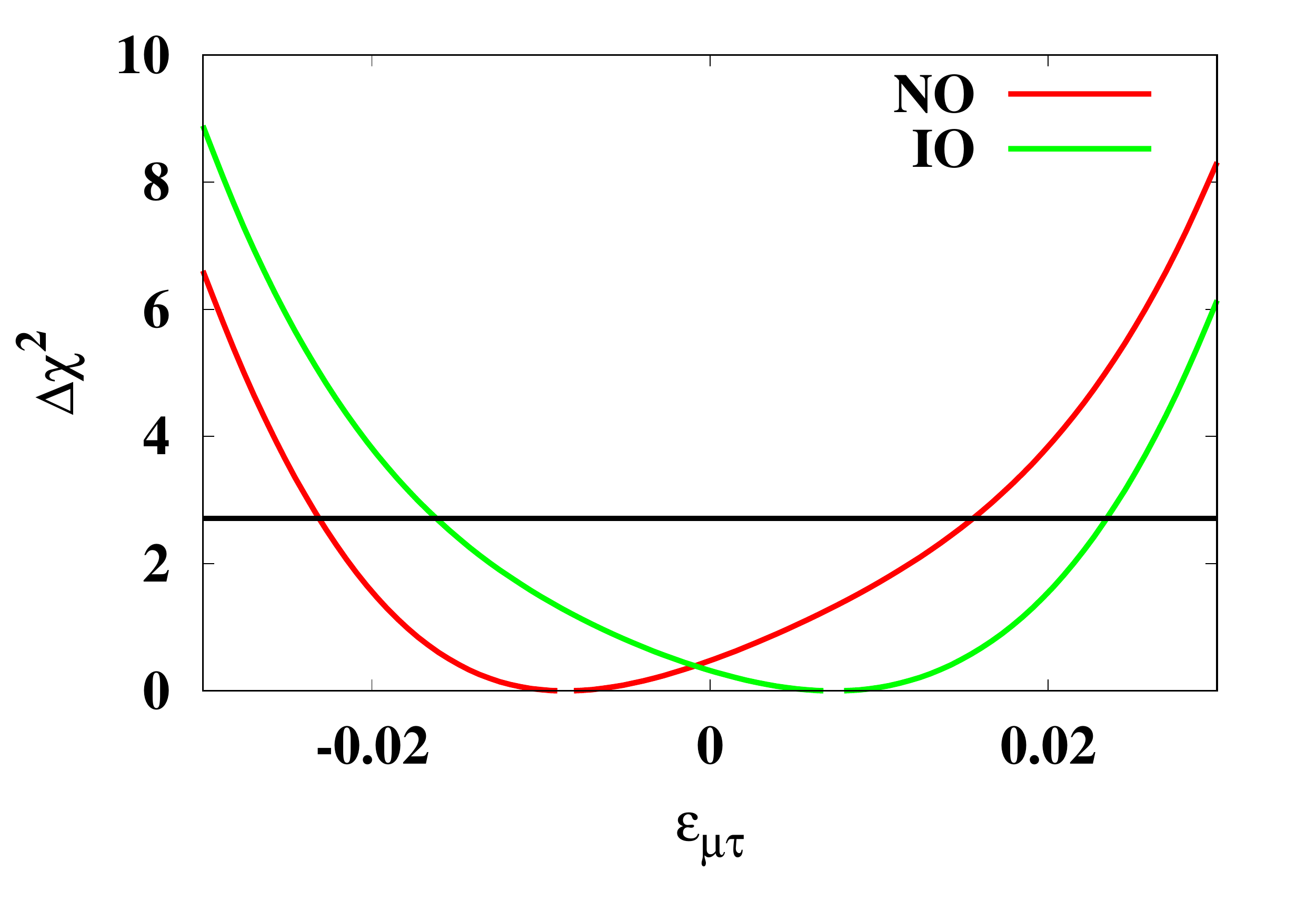}}
\put(240,140){\includegraphics[angle=-90,width=0.40\textwidth]{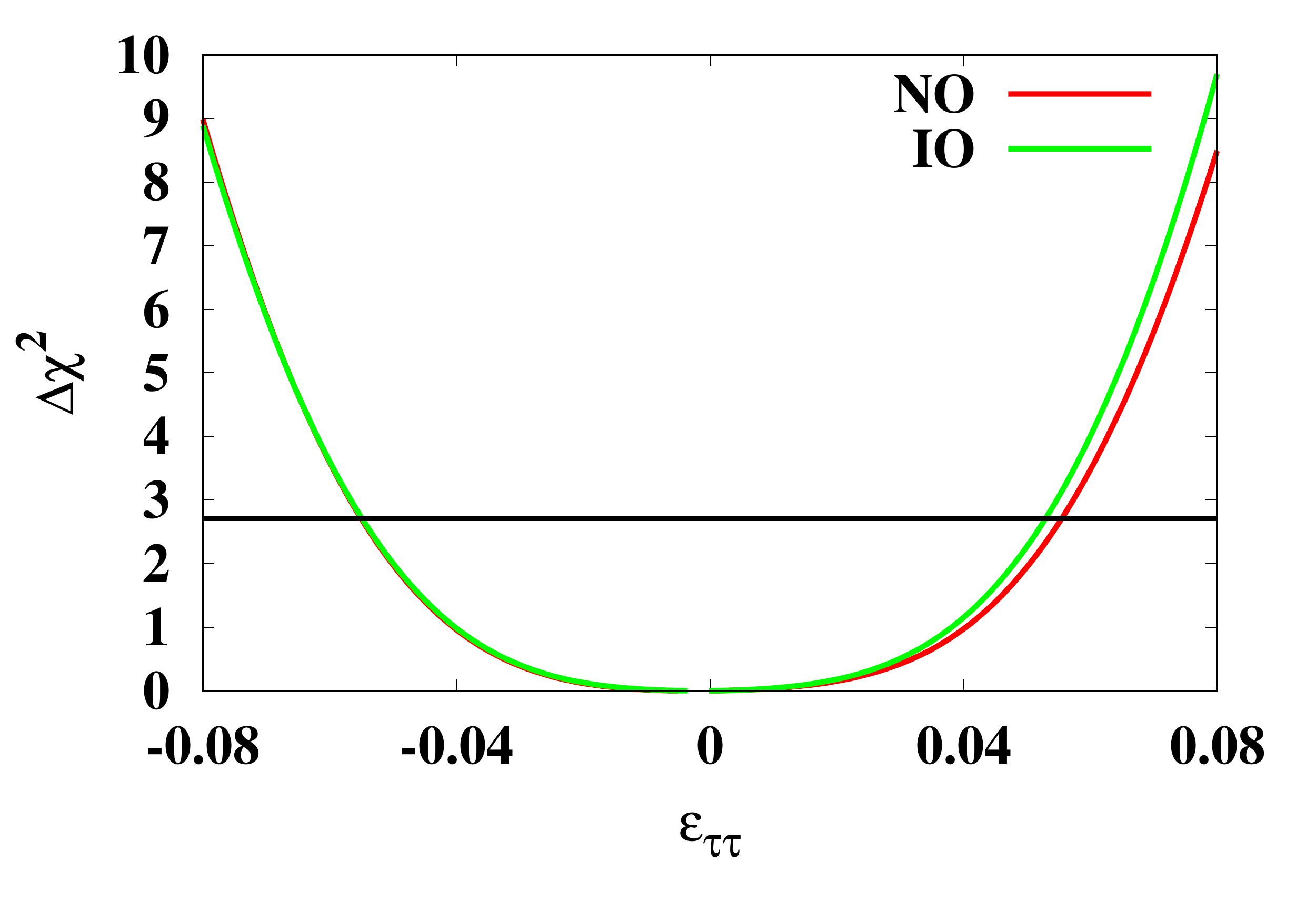}}
\end{picture}
\caption{\label{mutau} Values of $\Delta\chi^2$ for the cases of
  non-zero  $\epsilon_{\mu\tau}$ (left panel) and $\epsilon_{\tau\tau}$
  (right panel).  Other matter NSI parameters except for
    $\epsilon_{\mu\tau}$ (left panel) and $\epsilon_{\tau\tau}$ (right
  panel) are set to zero.
  Red (dark gray) and green (light gray) lines
  correspond to 
  normal and inverted neutrino mass hierarchy, respectively.} 
\end{figure}

Let us start with the $\mu\tau$ sector where the NSI parameters,
$\epsilon_{\mu\tau}$ and
$\epsilon^\prime\equiv\epsilon_{\tau\tau}-\epsilon_{\mu\mu}$, are
known to be severely constrained from experiments with atmospheric
neutrinos~\cite{Esmaili:2013fva,Salvado:2016uqu,Aartsen:2017xtt}. Firstly,
we take a single non-zero matter NSI 
parameter, $\epsilon_{\mu\tau}$ or $\epsilon_{\tau\tau}$, at a time
and perform minimization of $\chi^2$ given by Eq.~\eqref{eq:2:1} with
respect to all other variables (including $\Delta m_{31}^2$, 
$\sin^2{\theta_{23}}$ and other nuisance parameters discussed in
previous Section). We consider the cases
of normal (NO) and inverted (IO) neutrino mass ordering
independently. The results for
$\Delta\chi^2\equiv \chi^2-\chi^2_{min}$, where
$\chi_{min}^2$ is an absolute minimum of $\chi^2$ for each case, are shown in
Fig.~\ref{mutau}
for non-zero $\epsilon_{\mu\tau}$ (left panel) and $\epsilon_{\tau\tau}$ (right
panel). We find the following single parameter allowed ranges at 90\%~C.L.
\begin{gather}
  \label{mutau_r}
  -0.023<\epsilon_{\mu\tau}<0.016\; \text{(NO)}\,,\;\;\;
  -0.016<\epsilon_{\mu\tau}<0.023\; \text{(IO)}\,,\\
  \label{tautau_r}
  -0.055<\epsilon_{\tau\tau}<0.056\; \text{(NO)}\,,\;\;\;
  -0.055<\epsilon_{\tau\tau}<0.053\; \text{(IO)}\,.
\end{gather}
 Note that the obtained regions for
$\epsilon_{\mu\tau}$ are consistent\footnote{With our
   convention~\eqref{eq:1:4} the values of $\epsilon_{\alpha\beta}$
   differ by a factor of $r\equiv N_d/N_e\approx 3$ from those 
   used
   e.g. in~\cite{Aartsen:2017xtt,Salvado:2016uqu,Mitsuka:2011ty}. }  
with those $-0.020<\epsilon_{\mu\tau}<0.024$ (NO) obtained by IceCube
collaboration~\cite{Aartsen:2017xtt} in a similar single parameter
analysis using three years of their data with upward going track
events selection.  The regions~\eqref{mutau_r},~\eqref{tautau_r}
  are also close to the 
preliminary IceCube bounds $|\epsilon_{\mu\tau}|\lsim 0.17$ and
$|\epsilon_{\tau\tau}|\lsim 0.04$ (NO) from an analysis of DeepCore
data reported\footnote{Note that
  the analysis~\cite{ic_talk2} assumes presence of the
  complex phases in the flavour changing NSI parameters. For consistency,
  here we cite preliminary bounds for real valued parameters.}
in~\cite{ic_talk2}.

\begin{figure}[t]
\begin{picture}(300,130)(0,20)
\put(30,140){\includegraphics[angle=-90,width=0.40\textwidth]{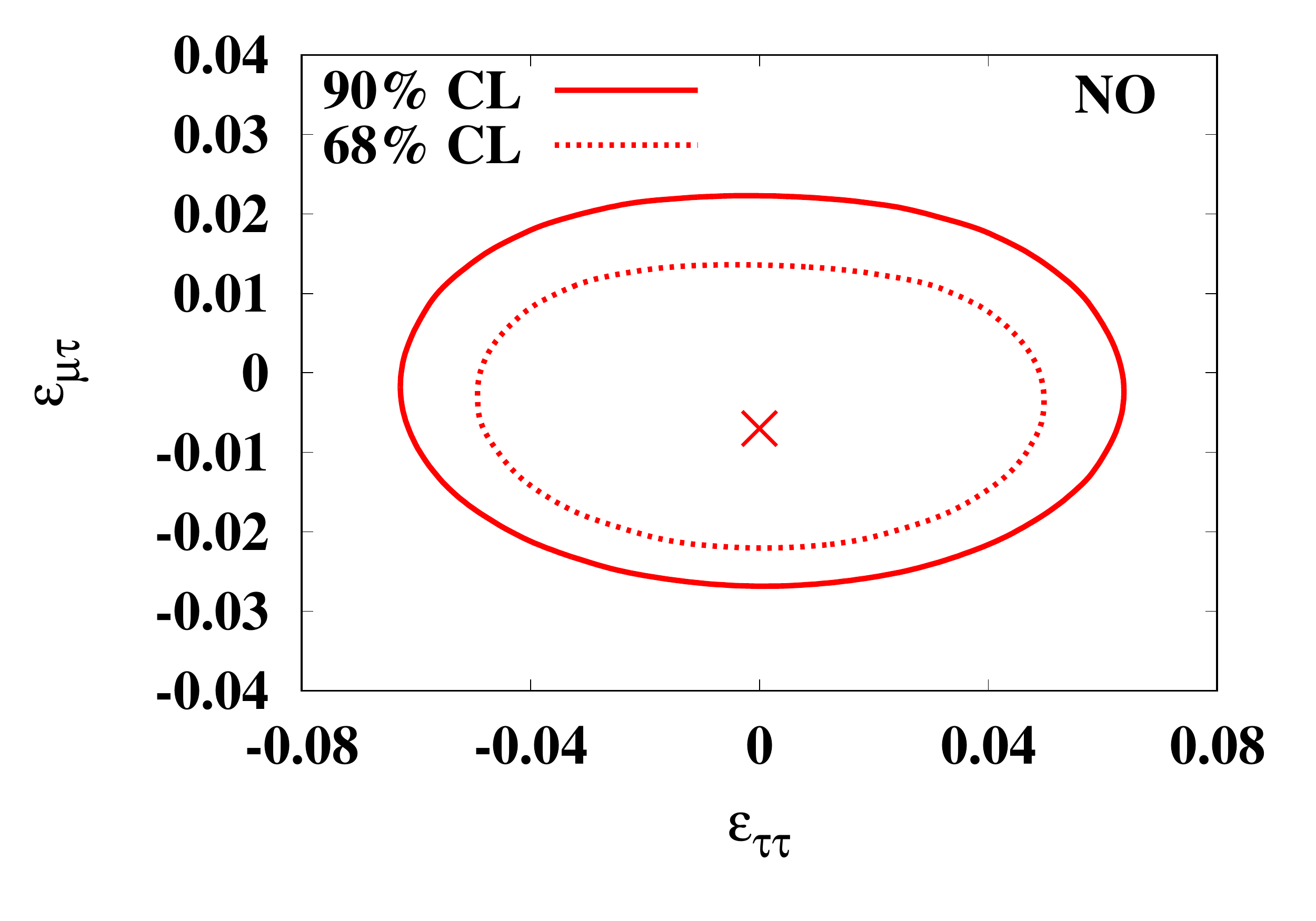}}
\put(240,140){\includegraphics[angle=-90,width=0.40\textwidth]{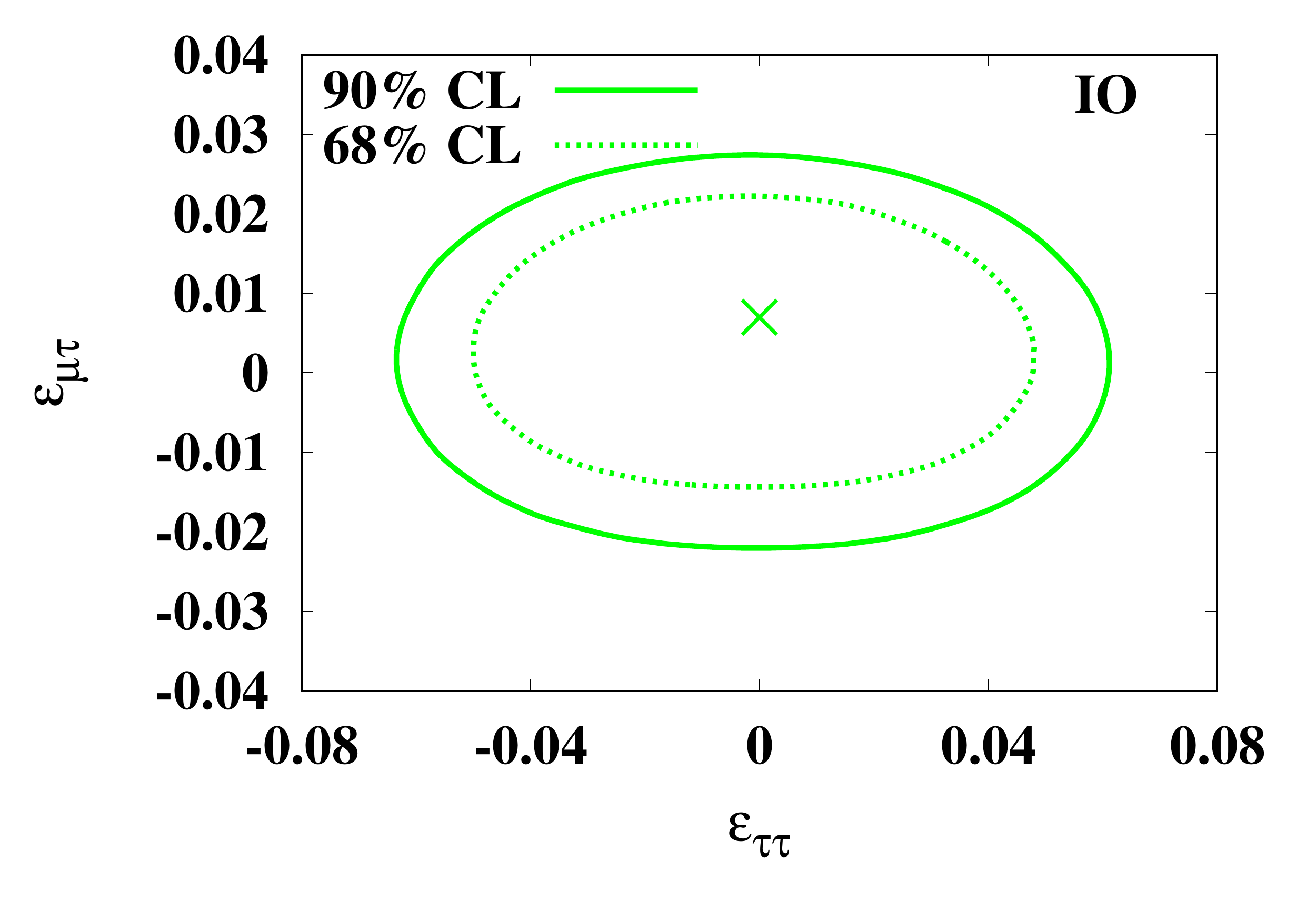}}
\end{picture}
\caption{\label{mutau_cor} Allowed regions for the parameters
  $\epsilon_{\mu\tau}$ and $\epsilon_{\tau\tau}$ for NO
  (left panel) and IO (right panel).  Other parameters
  $\epsilon_{\alpha\beta}$ of the matter NSI are set to   zero.}
\end{figure}

Next, we consider the NSI models in which both parameters in the
$\mu\tau$ sector, i.e. $\epsilon_{\mu\tau}$ and
$\epsilon_{\tau\tau}$. are not equal to zero. Corresponding allowed
regions on 
$(\epsilon_{\mu\tau},\epsilon_{\tau\tau})$ plane are presented in  
Fig.~\ref{mutau_cor} for NO (left panel) and IO (right panel). 
After marginalization with respect to each of these parameters we
obtain the following allowed ranges at 90\%~C.L.
\begin{gather}
  \label{mutau_marg_r}
  -0.027<\epsilon_{\mu\tau}<0.022\; \text{(NO)}\,,\;\;\;
  -0.022<\epsilon_{\mu\tau}<0.027\; \text{(IO)}\,,\\
  \label{tautau_marg_r}
  -0.063<\epsilon_{\tau\tau}<0.064\; \text{(NO)}\,,\;\;\;
  -0.064<\epsilon_{\tau\tau}<0.061\; \text{(IO)}\,,
\end{gather}
which are only slightly wider than those found in the single
parameter analysis and shown in~\eqref{mutau_r},\eqref{tautau_r}.  The
  marginalized bounds on  
$\epsilon_{\mu\tau}$ are weaker than the allowed range
$-0.018<\epsilon_{\mu\tau} < 0.016$~(NO), 90\%~C.I. obtained in 
Ref.~\cite{Salvado:2016uqu} from the analysis of a one
year high energy IceCube data sample~\cite{TheIceCube:2016oqi} and
than the earlier bound $-0.018<\epsilon_{\mu\tau}<0.017$ (NO) at
90\%~C.L. obtained in Ref.~\cite{Esmaili:2013fva} using the data from
79-string IceCube configuration and DeepCore.  At the same time the
marginalized 
bounds~\eqref{tautau_marg_r} on the parameter $\epsilon^\prime\equiv
\epsilon_{\tau\tau}$ are somewhat better than the allowed range
$-0.11<\epsilon^\prime<0.09$ at 90\%~C.L. from
Ref.~\cite{Esmaili:2013fva} and the constraint
$|\epsilon^\prime|<0.15$ obtained by Super-Kamiokande
experiment~\cite{Mitsuka:2011ty}. Comparable bound on
$\epsilon^\prime$ was obtained in Ref.~\cite{Salvado:2016uqu} from
a combination of high energy IceCube data and Super-Kamiokande results. 
Let us note that the effect of $\epsilon^\prime$ on neutrino propagation
decreases with neutrino energy (see e.g.~\cite{Esmaili:2013fva}) and
thus to probe this parameter one should rely on low energy part of the
atmospheric neutrino spectrum. 

\begin{figure}[t]
\begin{picture}(300,130)(0,20)
\put(30,140){\includegraphics[angle=-90,width=0.40\textwidth]{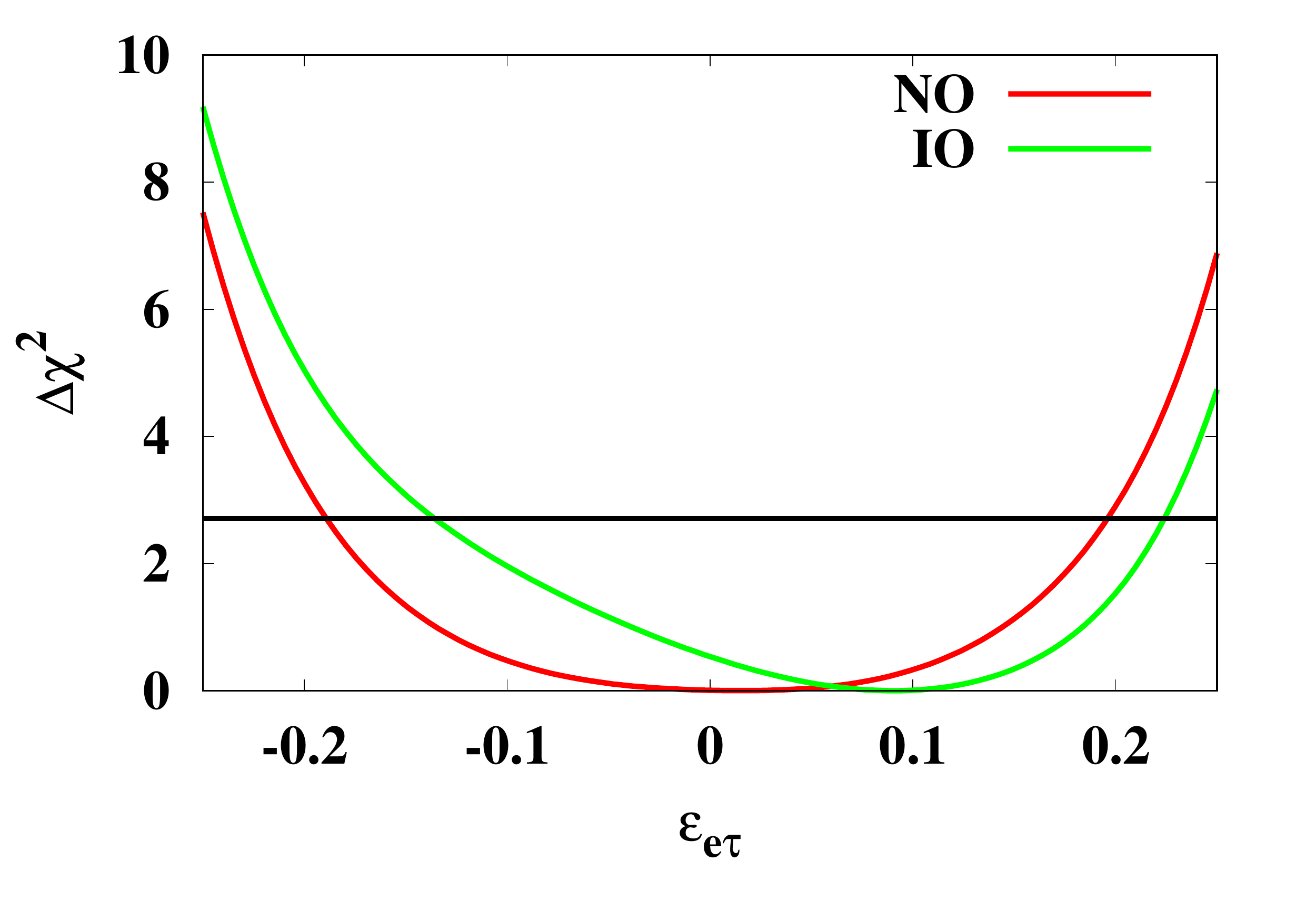}}
\put(240,140){\includegraphics[angle=-90,width=0.40\textwidth]{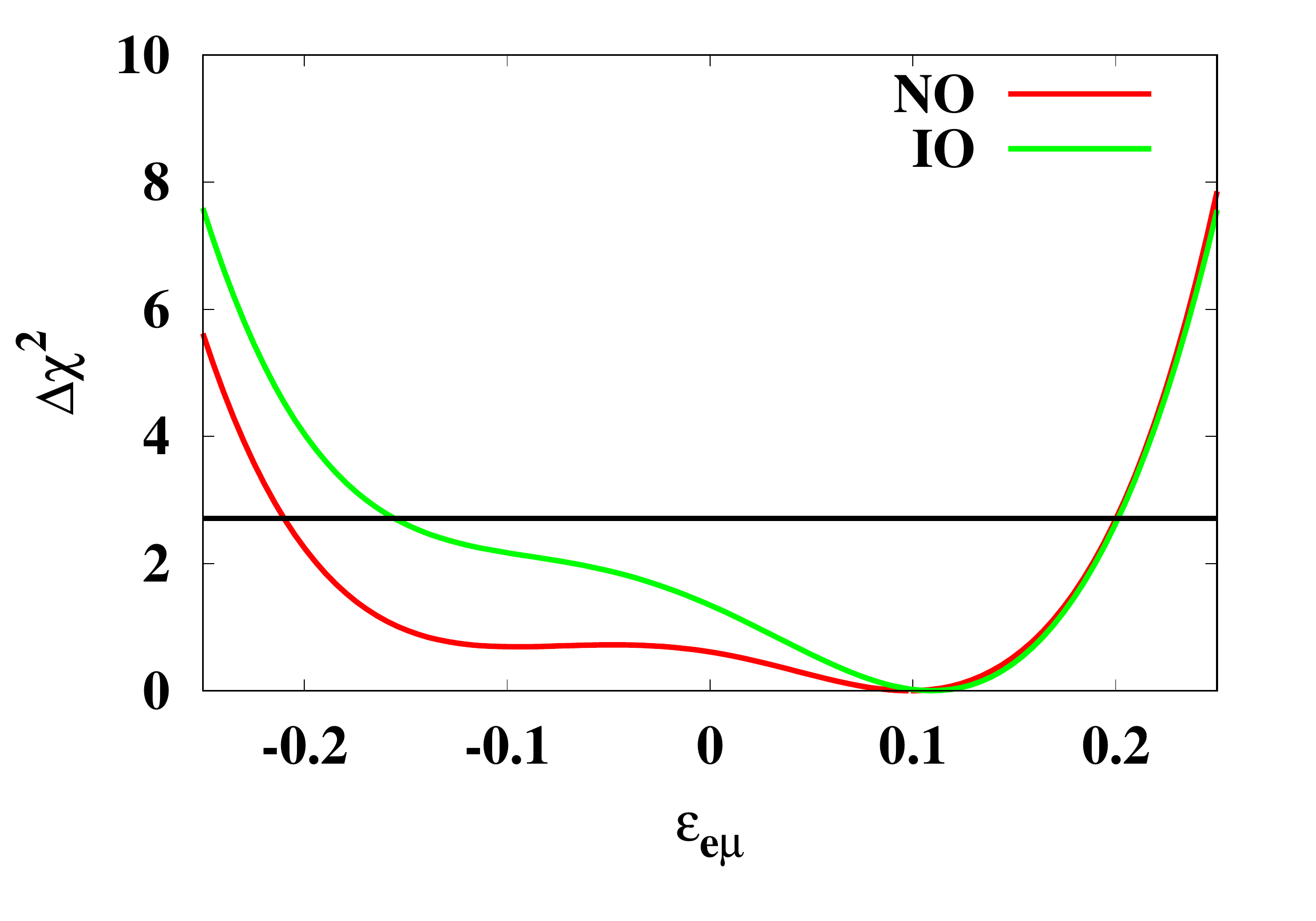}}
\end{picture}
\caption{\label{e} Values of $\Delta\chi^2$ for the cases of non-zero 
  $\epsilon_{e\tau}$ (left panel) and $\epsilon_{e\mu}$ (right
  panel).  Other NSI parameters except for $\epsilon_{e\tau}$
    (left panel) and $\epsilon_{e\mu}$ (right panel) are set to
    zero.  
   Red (dark gray) and green (light gray) lines correspond to NO and
   IO, 
  respectively.}
\end{figure}

\begin{figure}[t]
\begin{picture}(300,130)(0,20)
\put(30,140){\includegraphics[angle=-90,width=0.40\textwidth]{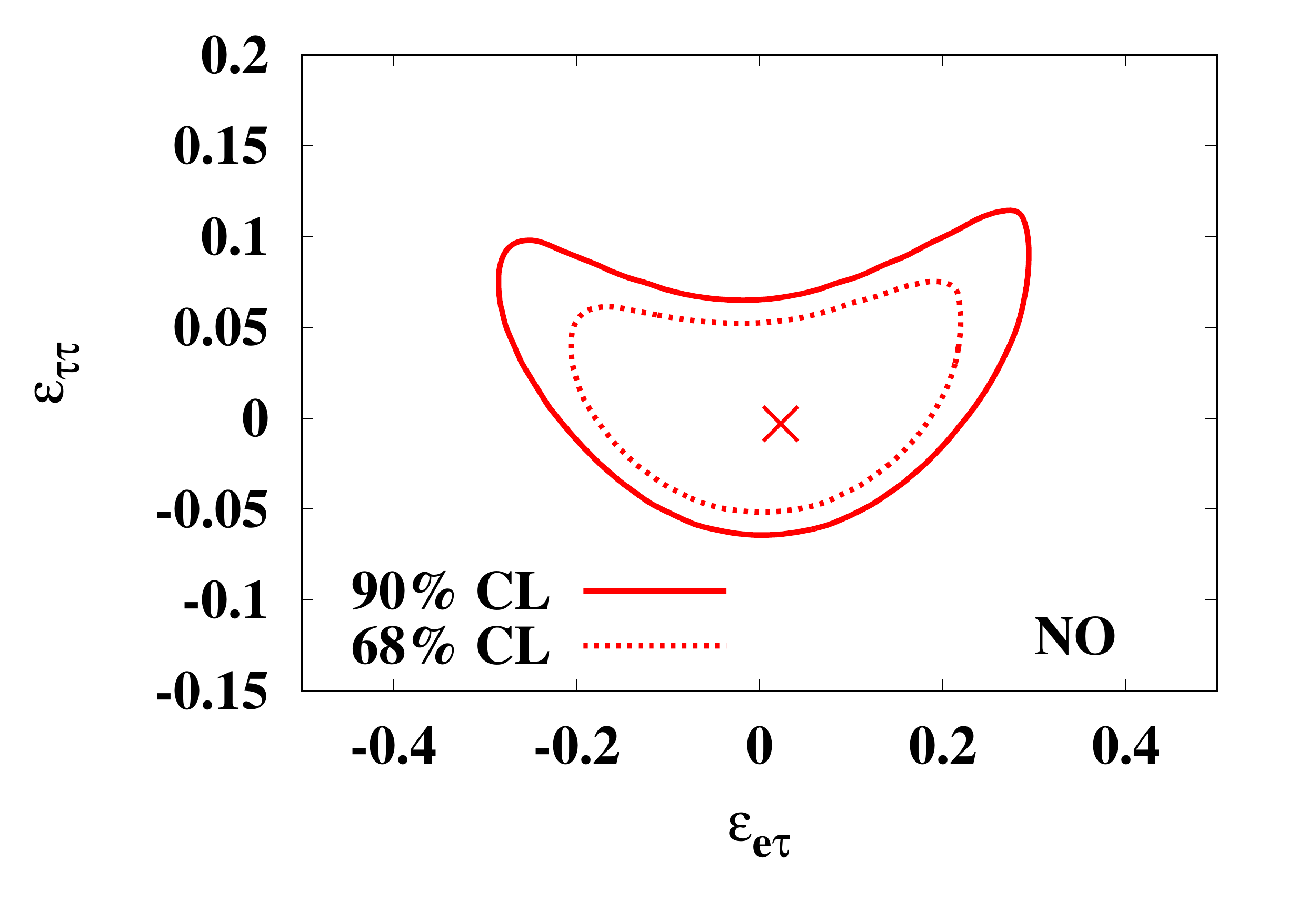}}
\put(240,140){\includegraphics[angle=-90,width=0.40\textwidth]{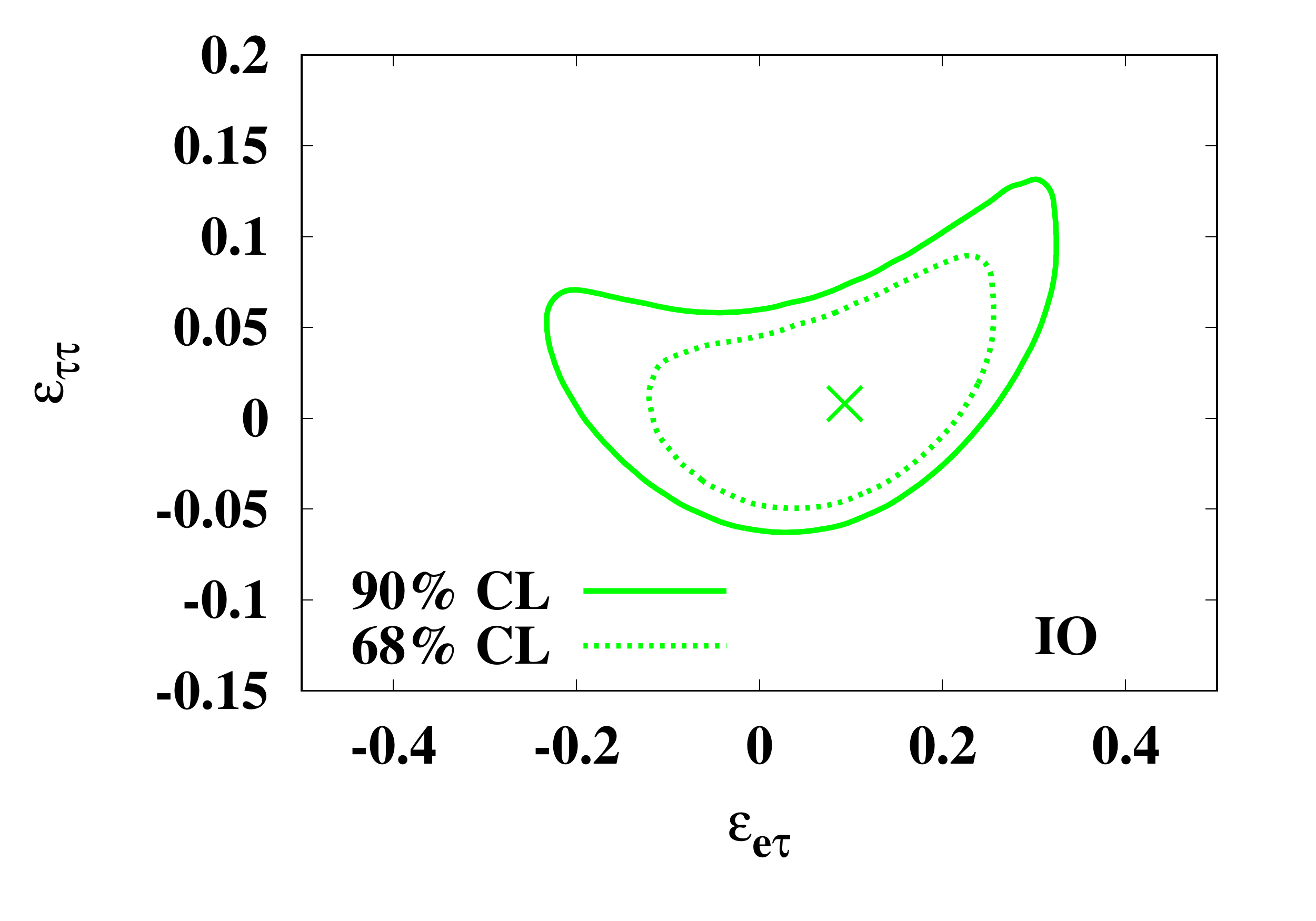}}
\end{picture}
\caption{\label{e_cor} Allowed regions for the parameters
  $\epsilon_{e\tau}$ and $\epsilon_{\tau\tau}$ for NO
  (left panel) and IO (right panel). Other parameters of the matter NSI including
  $\epsilon_{ee}$ are set to zero.}
\end{figure}

\begin{figure}[t]
\begin{picture}(300,130)(0,20)
\put(30,140){\includegraphics[angle=-90,width=0.40\textwidth]{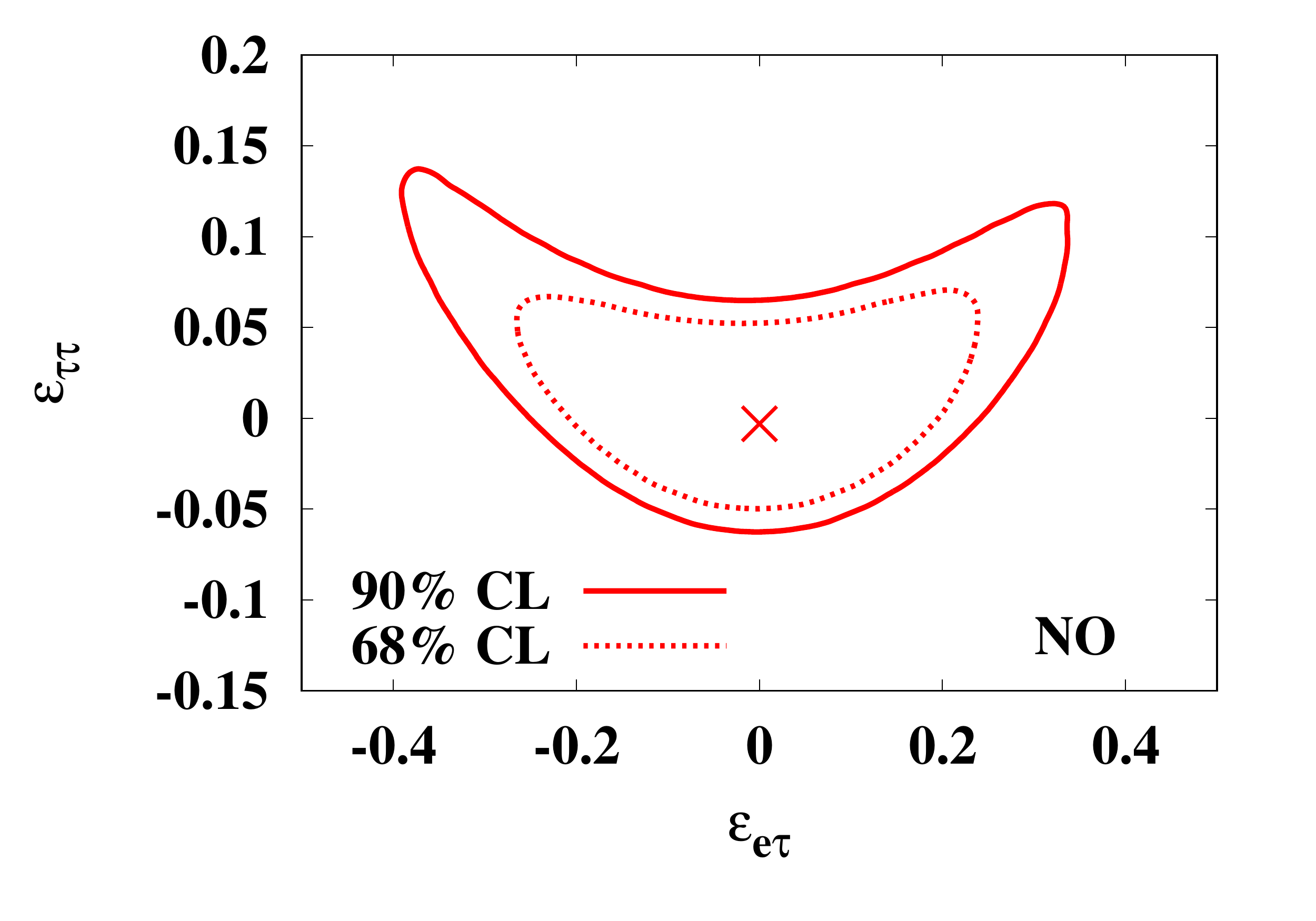}}
\put(240,140){\includegraphics[angle=-90,width=0.40\textwidth]{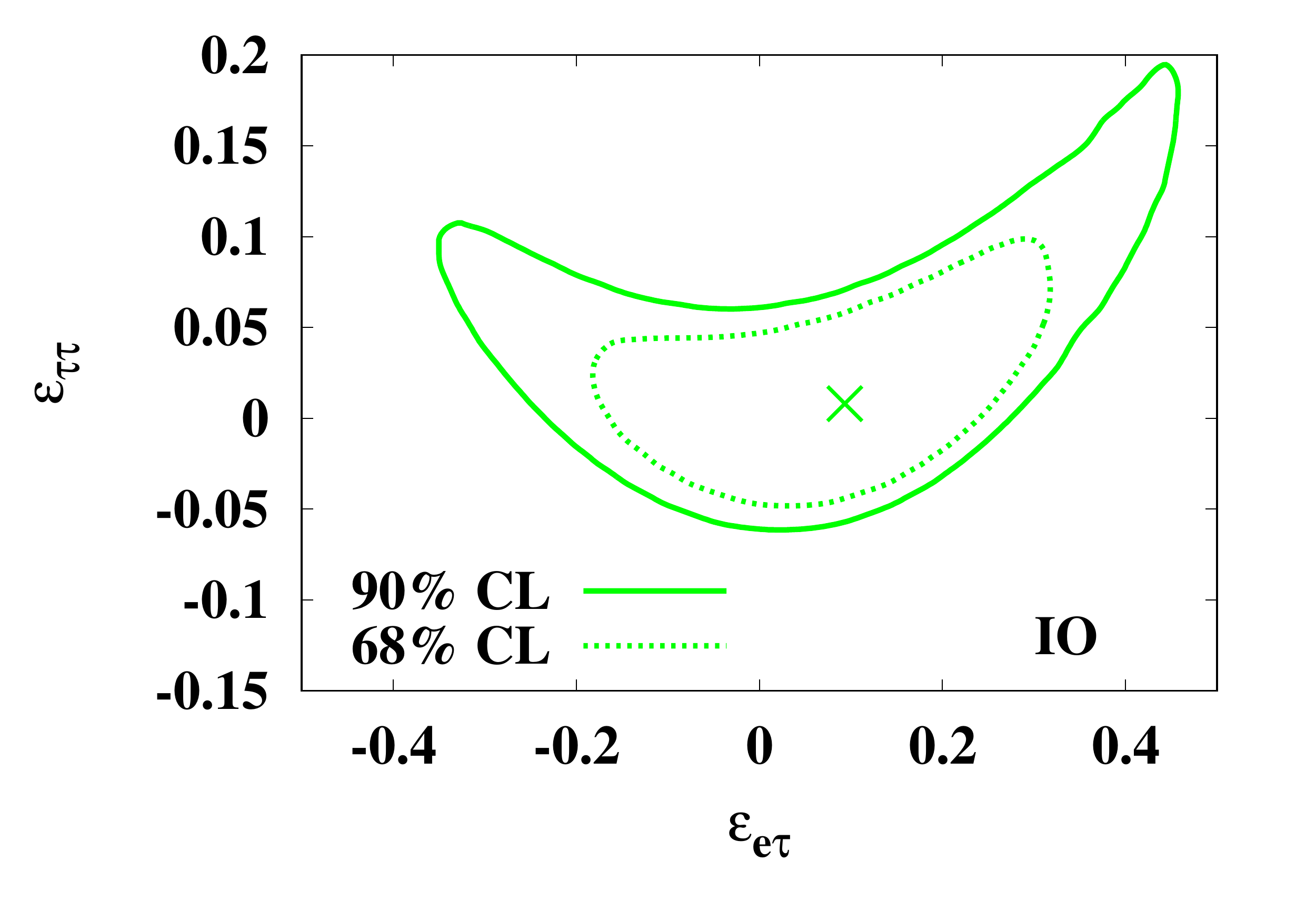}}
\end{picture}
\caption{\label{e02_cor} The same as in Fig.~\ref{e_cor} but for
  $\epsilon_{ee} = 0.2$.}
\end{figure}

\begin{figure}[t]
\begin{picture}(300,130)(0,20)
\put(30,140){\includegraphics[angle=-90,width=0.40\textwidth]{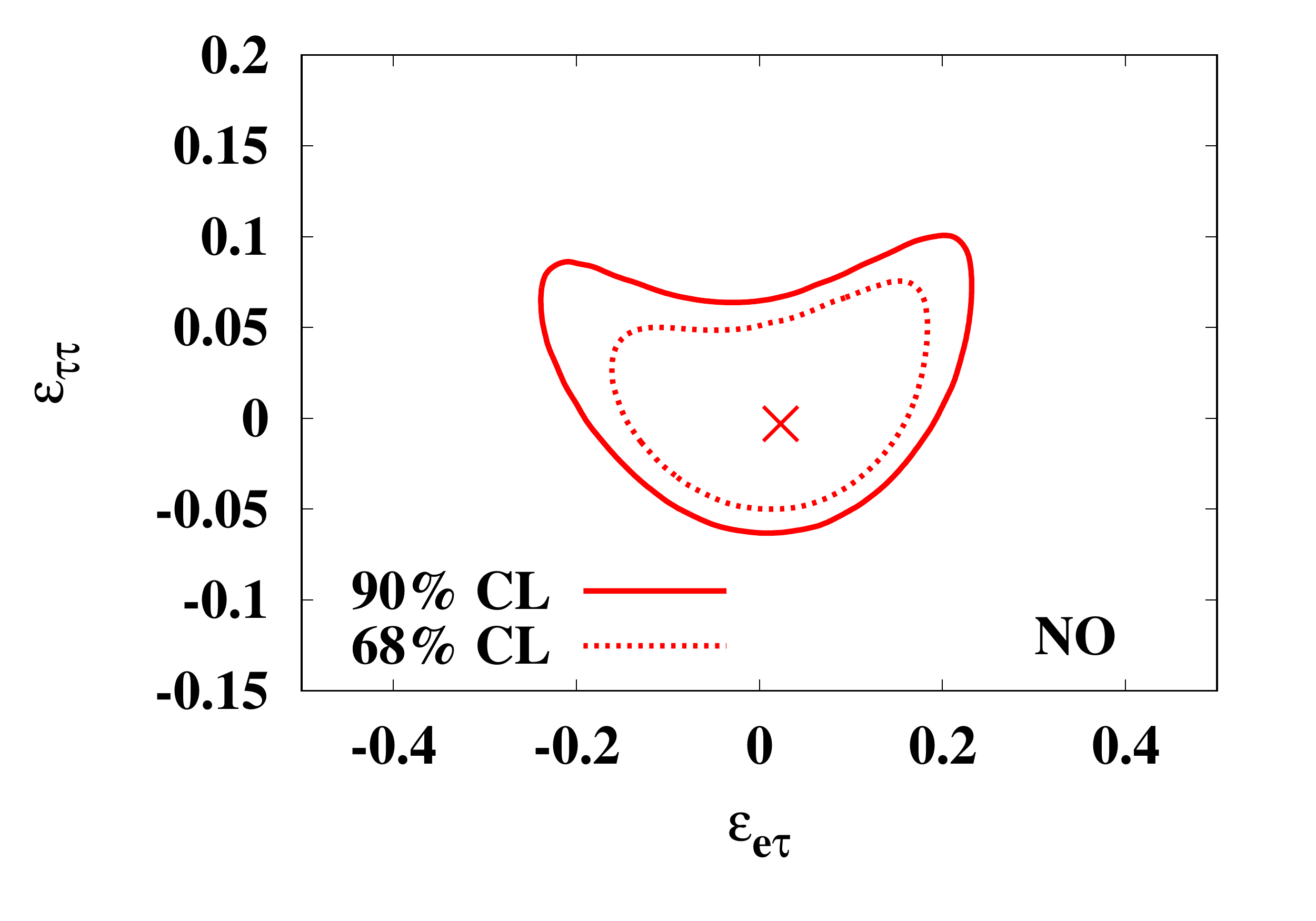}}
\put(240,140){\includegraphics[angle=-90,width=0.40\textwidth]{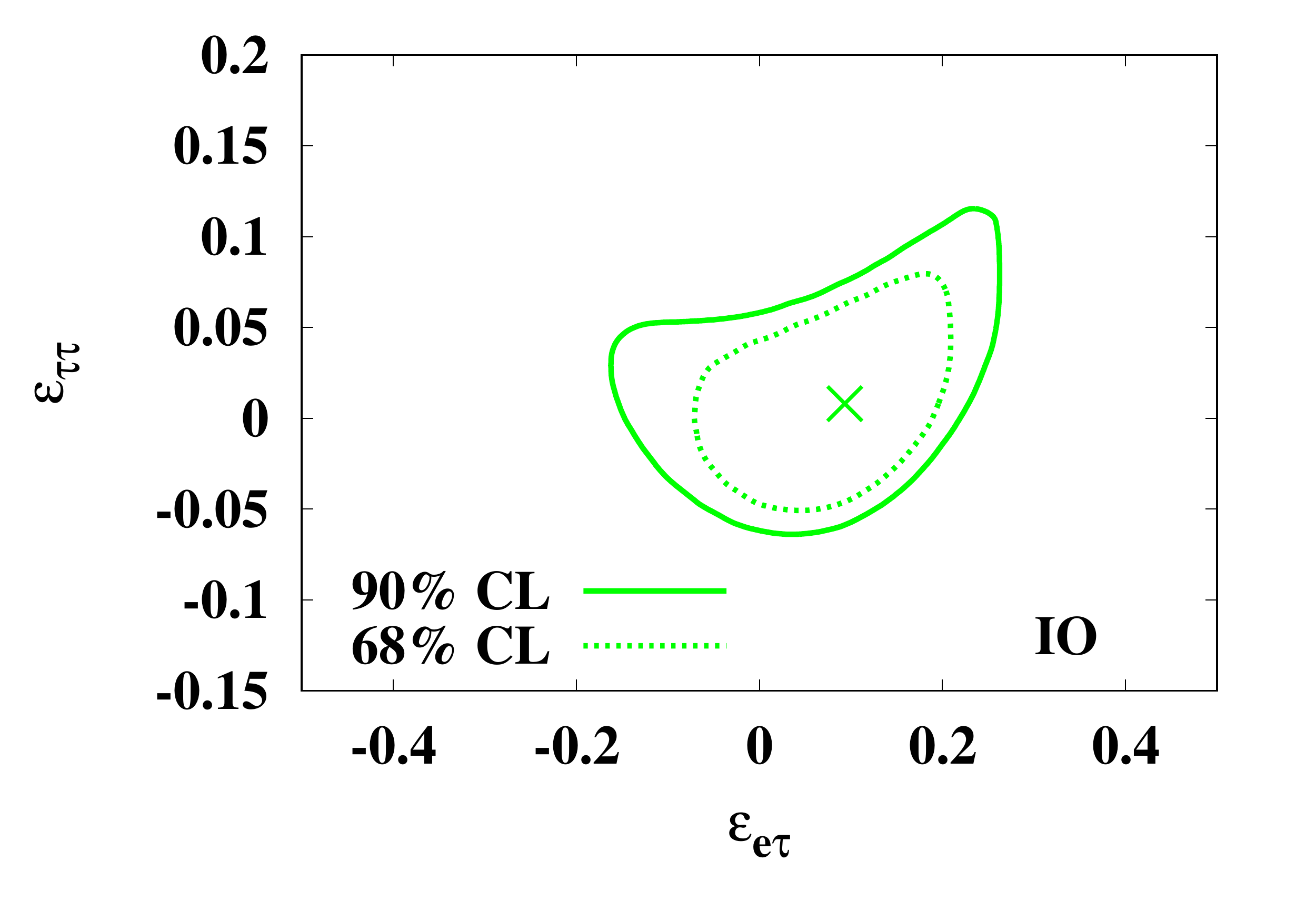}}
\end{picture}
\caption{\label{e-02_cor} The same as in Fig.~\ref{e_cor} but for
  $\epsilon_{ee} = -0.2$.}
\end{figure}

Now let us turn to the NSI models with non-zero parameters involving electron
neutrinos. In Fig.~\ref{e}
we show $\Delta\chi^2$ for NSI models with non-zero flavor changing
parameters  $\epsilon_{e\tau}$
(left panel) and $\epsilon_{e\mu}$ (right panel) taking a single
  non-zero NSI parameter at a time. Corresponding allowed
regions at 90\%~C.L. read
\begin{gather}
  \label{e_r}
  -0.21<\epsilon_{e\mu}<0.20\; \text{(NO)}\,,\;\;\;
  -0.16<\epsilon_{e\mu}<0.20\; \text{(IO)}\,,\\
  \label{e1_r}
  -0.19<\epsilon_{e\tau}<0.20\; \text{(NO)}\,,\;\;\;
  -0.14<\epsilon_{e\tau}<0.22\; \text{(IO)}\,.
\end{gather}
 The allowed single parameter bounds for $\epsilon_{e\mu}$ and
$\epsilon_{e\tau}$ are consistent with the preliminary results
$|\epsilon_{e\mu}|\lsim 0.16$ and $|\epsilon_{e\tau}|\lsim 0.2$ (NO)
reported by IceCube~\cite{ic_talk2}. Note that based on the behaviour
of $\Delta P_{\mu\mu}$ and $\Delta 
P_{\mu e}$ shown in Fig.~\ref{emu_osc} one can expect an improvement in the
allowed ranges on $\epsilon_{e\mu}$ with data on atmospheric neutrino
at high energies.
It is well known that the effect
of $\epsilon_{e\tau}$ on neutrino transition probabilities strongly
depends on the values of other matter NSI parameters. Following
the analysis~\cite{Mitsuka:2011ty} and earlier 
studies~\cite{Friedland:2004ah,Friedland:2005vy} let us consider the
NSI models with non-zero parameters in $e\tau$ sector,
i.e. $\epsilon_{ee}, \epsilon_{e\tau}$ and $\epsilon_{\tau\tau}$.
In Fig.~\ref{e_cor}
we show allowed regions for $\epsilon_{e\tau}$ and
$\epsilon_{\tau\tau}$ within these models assuming
$\epsilon_{ee}=0$. The parabolic 
form of the allowed region comes from the approximate relation
$\epsilon_{\tau\tau} \sim
\frac{|\epsilon_{e\tau}|^2}{1+\epsilon_{ee}}$, which should be
satisfied for consistency with high energy part of the atmospheric
neutrino spectrum, see
\cite{Friedland:2004ah,Friedland:2005vy}.  Similar allowed regions
from preliminary analysis of tree-year DeepCore data were presented
in~\cite{ic_poster}. 
The sensitivity of the considered dataset to $\epsilon_{ee}$ is
relatively weak, but still 
its non-zero value may affect the bounds on $\epsilon_{e\tau}$ and
$\epsilon_{\tau\tau}$ in this class of NSI models through the above
relation. As an illustration, in Figs.~\ref{e02_cor} and~\ref{e-02_cor}
are shown the allowed regions in
$(\epsilon_{\tau\tau},\epsilon_{e\tau})$ plane assuming
$\epsilon_{ee}=0.2$ and~$-0.2$, respectively. We see the dependence
of the bounds for $\epsilon_{e\tau}$ on the assumption about the value of
$\epsilon_{ee}$. The same is valid for allowed regions for
$\epsilon_{\tau\tau}$ which can be considerably modified as compared
to those given by Eq.~\eqref{tautau_marg_r}  obtained for the NSI
models with other $\epsilon_{\alpha\beta}=0$. The found allowed
regions for $\epsilon_{e\tau}$ and 
$\epsilon_{\tau\tau}$ are compatible to the latest constraints on
these parameters obtained with the Super-Kamiokande data in
Ref.~\cite{Fukasawa:2015jaa} using somewhat different anzatz for the
parameter space.

\begin{figure}[t]
\begin{picture}(300,260)(0,20)
\put(30,140){\includegraphics[angle=-90,width=0.40\textwidth]{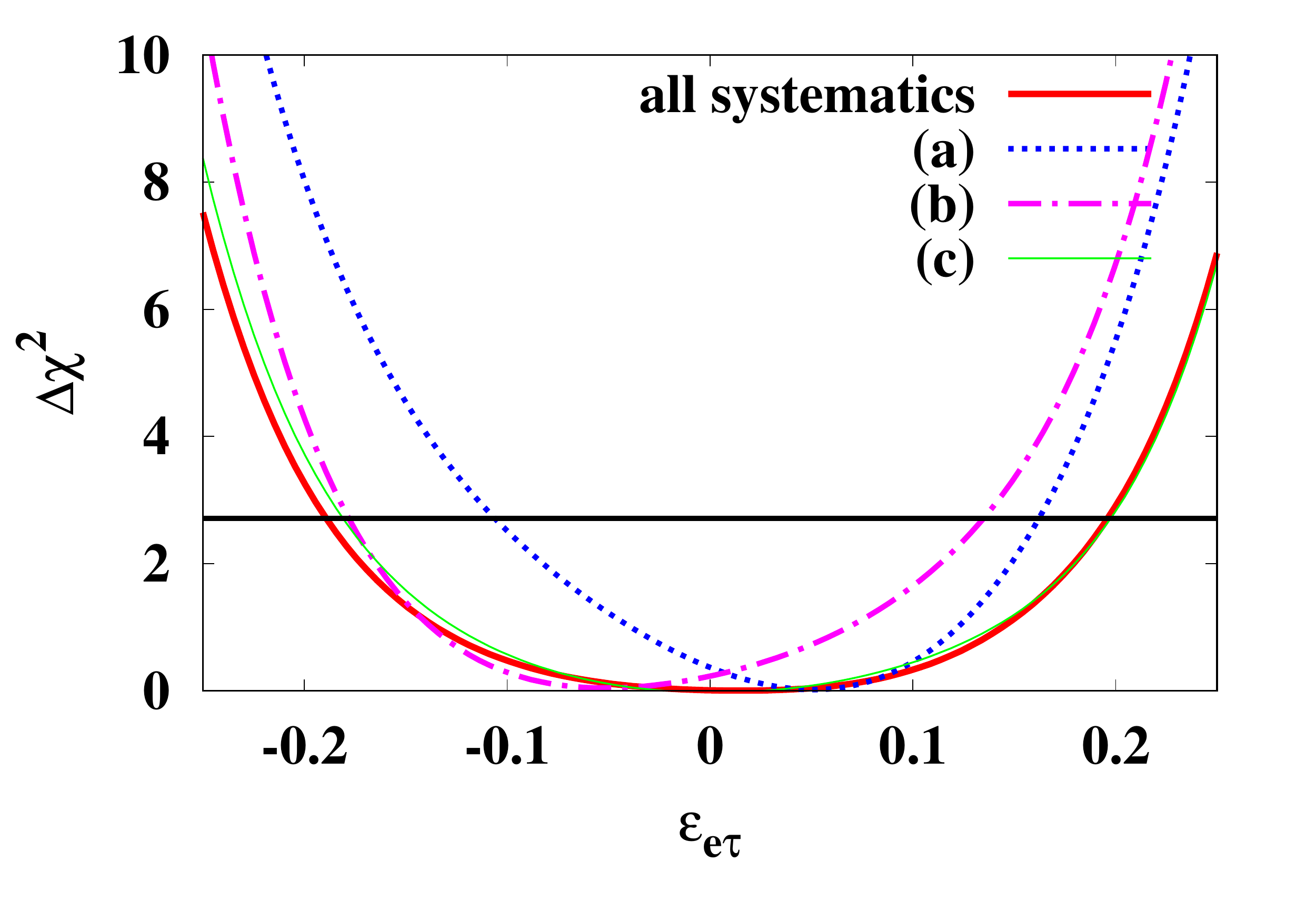}}
\put(240,140){\includegraphics[angle=-90,width=0.40\textwidth]{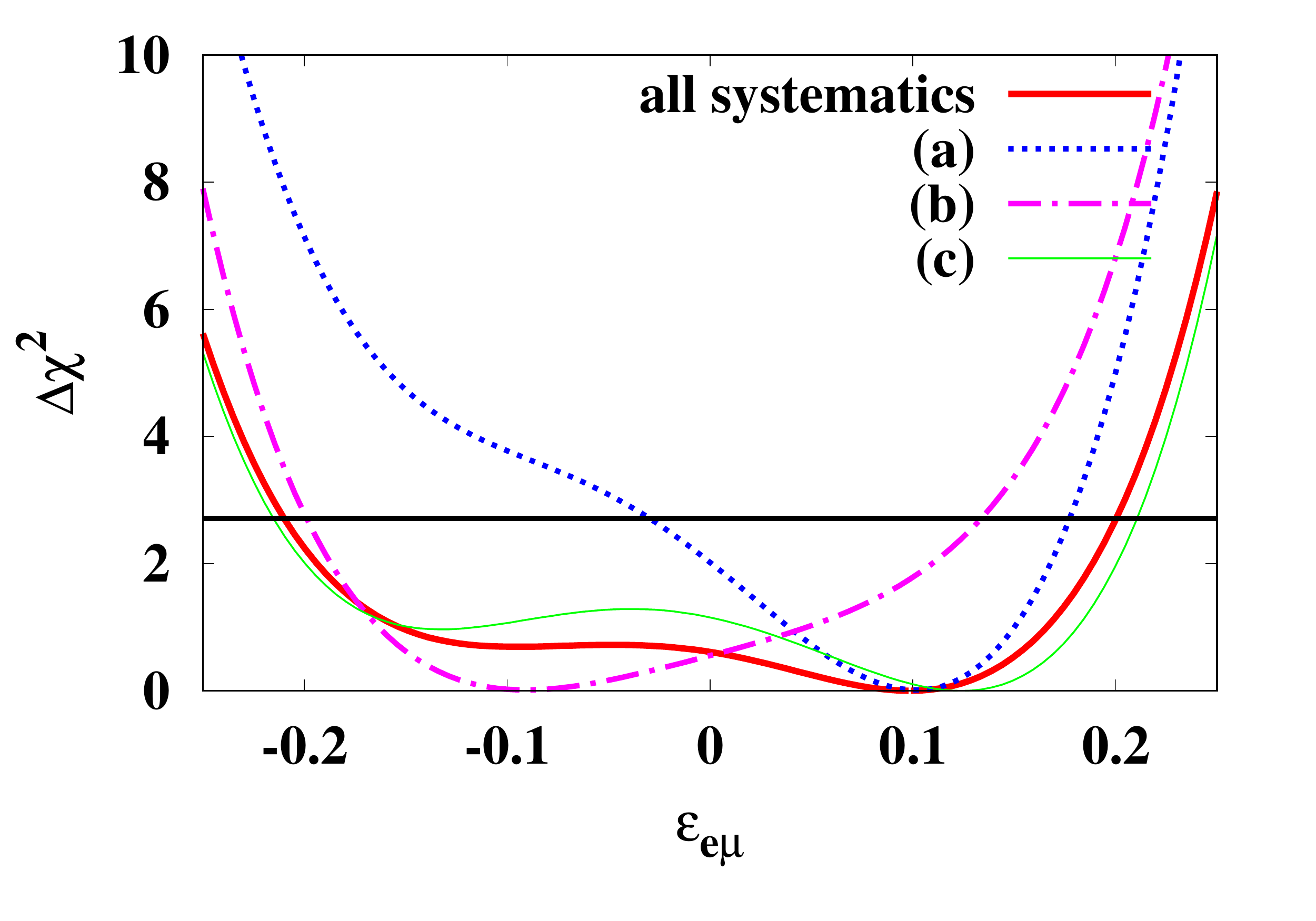}}
\put(30,280){\includegraphics[angle=-90,width=0.40\textwidth]{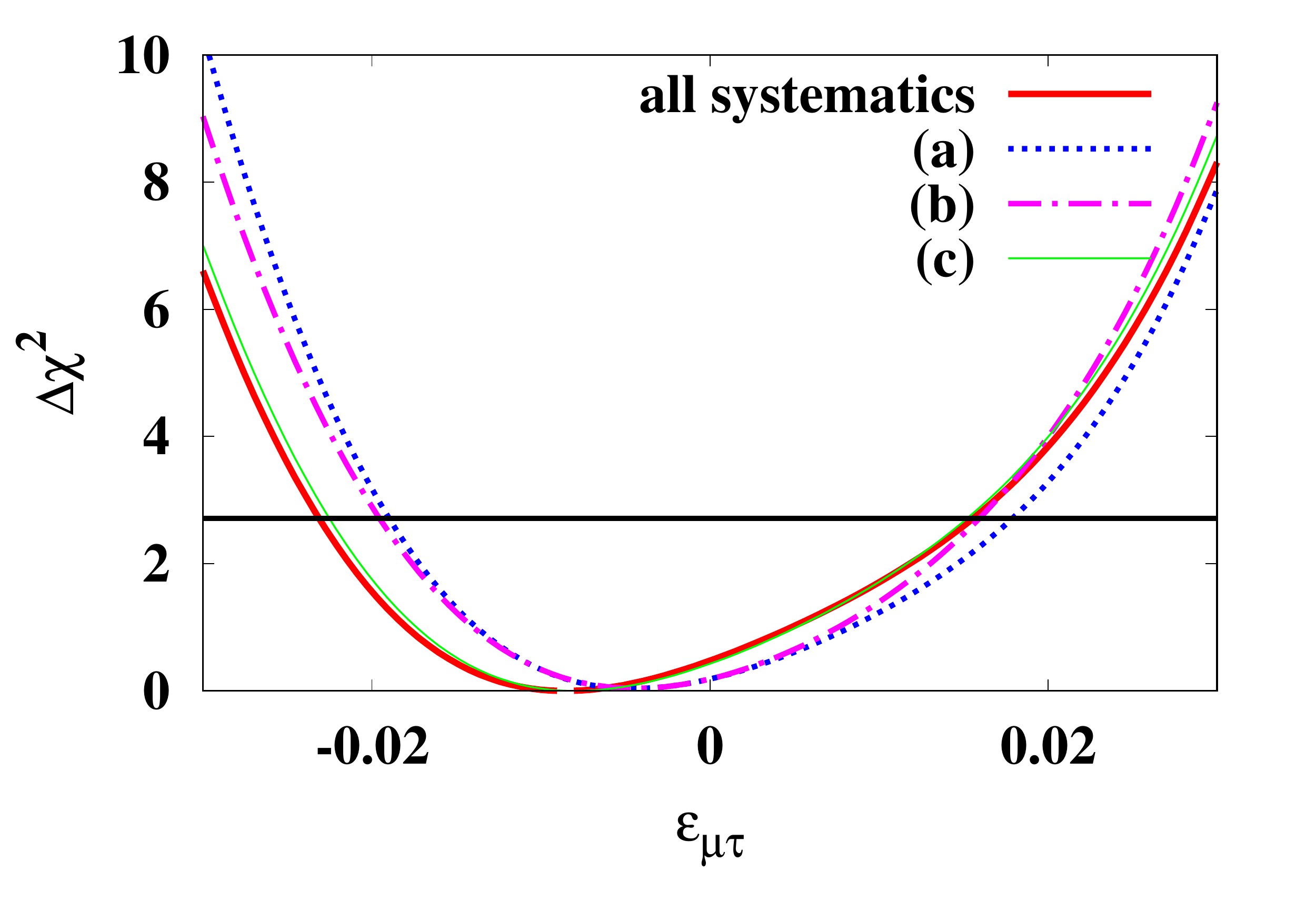}}
\put(240,280){\includegraphics[angle=-90,width=0.40\textwidth]{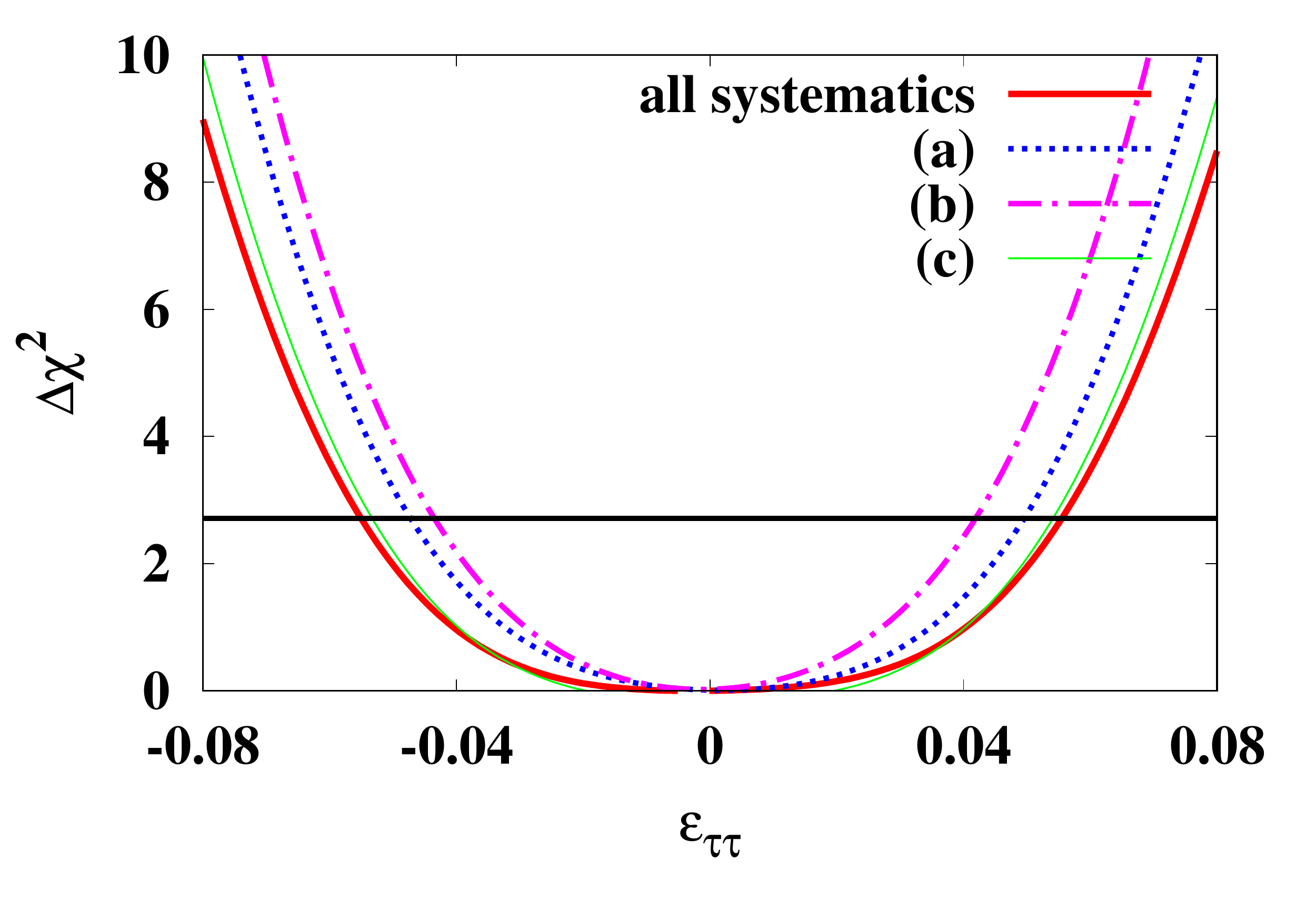}}
\end{picture}
\caption{\label{syst_effect} Values of $\Delta \chi^2$ (NO) for the
  NSI models with non-zero values of $\epsilon_{\mu\tau}$ (upper left),
  $\epsilon_{\tau\tau}$   (upper right), $\epsilon_{e\tau}$ (lower
  left) and $\epsilon_{e\mu}$ 
  (lower right) obtained in the analyses with all systematic
  uncertainties included (red thick solid line) in comparison with
  those obtained with (a) fixed nuisance parameters related to
  atmospheric neutrino flux (blue dotted line), (b) fixed nuisance
  parameters related to 
  the experimental uncertainties (magenta dashed-dotted line) and (c)
  fixed nuisance parameters related to normalization of different
  contributions to expected number of events (green thin solid line).} 
\end{figure}

Finally we study impact of different nuisance parameters
  discussed in Section~2 on our
  results. Firstly, in
Fig.~\ref{syst_effect} 
we present $\Delta\chi^2$ obtained in the single-parameter
analyses (for NSI models with non-zero $\epsilon_{\mu\tau}$,
$\epsilon_{\tau\tau}$, $\epsilon_{e\tau}$ or  $\epsilon_{e\mu}$ and
assuming normal neutrino 
mass ordering) for the case with all systematic uncertainties included
(which is our default analysis) in comparison with those obtained with
(a) fixed nuisance parameters 
related directly to the uncertainties in atmospheric neutrino flux,
i.e. spectral index and uncertainties in the hadron production in
atmosphere, (b) fixed nuisance parameters related to the experimental
systematic uncertainties and (c) fixed nuisance parameters which
determine normalizations of different contributions to expected number
of events including those related to the neutrino nucleon cross section
and atmospheric muon background. For illustration we present results
for the case
of normal mass ordering. One observes that the most important
systematic uncertainties are the experimental ones and those related to
atmospheric neutrino flux, specifically for the NSI models with non-zero 
$\epsilon_{e\tau}$ and $\epsilon_{\mu\tau}$.
We also study impact of neutrino oscillation parameters,
i.e. $\sin^{2}{\theta_{23}}$ and $\Delta m^2_{31}$. As an example, we
fix them to their optimal values without NSI for the case of normal
mass ordering (see Section~2). Corresponding allowed ranges for
$\epsilon_{\mu\tau}, \epsilon_{\tau\tau}, 
\epsilon_{e\tau}, \epsilon_{e\mu}$ (a single non-zero NSI parameter at
a time) becomes 
\begin{gather}
  -0.19 < \epsilon_{\mu\tau} < 0.13\,,\;\;\;
  -0.052 < \epsilon_{\tau\tau} < 0.051\,,\\
  -0.17 < \epsilon_{e\tau} < 0.16\,,\;\;\;
  -0.20 < \epsilon_{e\mu} < 0.18\,.
\end{gather}
By comparison
with~\eqref{mutau_r},\eqref{tautau_r} and~\eqref{e_r},\eqref{e1_r} we
see that the oscillation parameters affect very little these bounds.

\section{Conclusions}

Let us summarize results of our study. Here we used the three-year
IceCube DeepCore data sample~\cite{ic_data} of low energy atmospheric 
neutrinos to constrain the parameters of non-standard neutrino
interactions in propagation. This data sample contains track-like as
well as cascade-like events which makes it sensitive not mainly to muon but
also to electron neutrino flux at the detector level. Using this
dataset we found the bounds on several $\epsilon_{\alpha\beta}$ under
different model assumptions. In particular, we presented allowed
regions 
for the matter NSI parameters, assuming 1) single non-zero parameter
at a time, 2) non-zero parameters in $\mu\tau$ sector and 3) non-zero
$\epsilon_{e\tau}$ and $\epsilon_{\tau\tau}$ for several fixed values
of $\epsilon_{ee}$. The bounds on $\epsilon_{\mu\tau}$ and
$\epsilon_{\tau\tau}$ were found to be close to
those~\cite{Esmaili:2013fva,Fukasawa:2015jaa,Salvado:2016uqu,Aartsen:2017xtt}
extracted from different data samples of the IceCube/DeepCore as well
as from Super-Kamiokande results. Obtained single-parameter
bounds on $\epsilon_{\mu\tau}$, $\epsilon_{\tau\tau}$, $\epsilon_{e\tau}$ and
$\epsilon_{e\mu}$ are consistent with the preliminary IceCube
results\cite{ic_poster,ic_talk1,ic_talk2} from a similar analysis of
three-year DeepCore data. We studied the impact of different sources
of systematic uncertainties and found that the obtained bounds are
mainly stable with respect to the nuisance parameters of the
analysis. The main errors in the
bounds results from uncertainties in the atmospheric neutrino flux and 
from experimental uncertainties. They are specifically important for
single parameter bounds on $\epsilon_{e\tau}$ and $\epsilon_{e\mu}$.
Note that from energy dependence of neutrino transition probabilities we can
expect an improvement in the bound on $\epsilon_{e\mu}$ using studies
with high energy part of atmospheric neutrino spectrum.

In the main text we made a comparison mainly with the results obtained 
from oscillation experiments with atmospheric neutrinos. The obtained
bounds on $\epsilon_{\mu\tau}$,   $\epsilon_{\tau\tau}$,
$\epsilon_{e\tau}$ and $\epsilon_{e\mu}$ are also consistent with the
results of global analysis of neutrino oscillation 
experiments~\cite{Esteban:2018ppq} (see
also~\cite{GonzalezGarcia:2011my,Gonzalez-Garcia:2013usa,Gonzalez-Garcia:2015qrr}
for earlier studies) obtained under assumptions that the matter NSI
are given by neutrino interactions with quarks only and that their
flavour structure is independent of the quark flavour. Further insight
on possible strength of NSI can be obtained by combining with the
results of scattering
experiments~\cite{Coloma:2017egw,Coloma:2019mbs} (see
also~\cite{Barranco:2007ej,Biggio:2009nt}). Here we were 
interested in the impact of the low energy IceCube DeepCore
data~\cite{ic_data} solely and leaved a combined analysis for future
study.

\vskip 0.3cm
I am grateful to Philipp Eller for helpful correspondence.
The work was supported by the RSF grant 17-12-01547.

\bibliographystyle{unsrt}

\end{document}